\documentclass[linenumbers,twocolumn]{aa}

\usepackage{longtable}
\usepackage{graphicx}
\usepackage{graphics}
\usepackage{color}
\usepackage{epsfig}
\usepackage{float}
\usepackage{url}
\usepackage{hyperref} 
\usepackage{breakurl}
\usepackage{scalerel}
\usepackage[titletoc,title]{appendix}
\usepackage{relsize}
\usepackage{bm}
\usepackage{natbib}
\usepackage{orcidlink}
\usepackage[switch]{lineno} 
\usepackage{soul}

\begin{document}
\titlerunning{CO observation with IRAM 30m of 23 galaxies}
\authorrunning{Y. Jiang et al.}

\title{CO-CHANGES II: spatially resolved IRAM 30\MakeLowercase{M} CO line observations of 23 nearby edge-on spiral galaxies}

\author{Yan Jiang\inst{1,2}\orcidlink{0009-0003-3907-5077}
\and
Jiang-Tao Li\inst{1}\orcidlink{0000-0001-6239-3821}\thanks{Corresponding author: pandataotao@gmail.com}
\and
Qing-Hua Tan\inst{1}\orcidlink{0000-0003-3032-0948}
\and
Li Ji\inst{1,3}
\and
Joel N. Bregman\inst{4}
\and
Q. Daniel Wang\inst{5}\orcidlink{0000-0002-9279-4041}
\and
Jian-Fa Wang\inst{1,2}\orcidlink{0000-0001-6693-0743}
\and
Li-Yuan Lu\inst{6,1}\orcidlink{0000-0002-3286-5346}
\and
Xue-Jian Jiang\inst{7}\orcidlink{0000-0002-8899-4673}
}

\institute{Purple Mountain Observatory, Chinese Academy of Sciences, 10 Yuanhua Road, Nanjing 210023, China
\and
School of Astronomy and Space Sciences, University of Science and Technology of China, Hefei 230026, China
\and
Key Laboratory of Dark Matter and Space Astronomy, CAS, Nanjing 210023, China
\and
Department of Astronomy, University of Michigan, 311 West Hall, 1085 S. University Ave, Ann Arbor, MI, 48109-1107, U.S.A.
\and
Department of Astronomy, University of Massachusetts, Amherst, MA 01003, U.S.A.
\and
Department of Astronomy, Xiamen University, 422 Siming South Road, Xiamen 361005, China
\and
Research Center for Intelligent Computing Platforms, Zhejiang Laboratory, Hangzhou 311100, China
}

\abstract
{
Molecular gas, serving as the fuel for star formation, and its relationship with atomic gas are crucial for understanding how galaxies regulate their star forming (SF) activities. 
}
{
We conduct IRAM 30m observations of 23 nearby spiral galaxies as part of the CHANG-ES (Continuum Halos in Nearby Galaxies – an EVLA Survey) project to investigate the distribution of molecular gas and the Kennicutt–Schmidt SF law in these galaxies. By combining these results with atomic gas masses studied in previous work, we aim to investigate the scaling relations that connect the molecular and atomic gas masses with stellar masses and the baryonic Tully-Fisher relation.

}
{Based on spatially resolved observations of the ${\rm ^{12}CO}$ $J=1-0$, ${\rm ^{13}CO}$ $J=1-0$, and ${\rm ^{12}CO}$ $J=2-1$ molecular lines, we calculate the total molecular gas masses, the ratios between different CO lines, and derive some key physical parameters such as the temperature and optical depth of the molecular gas. 
}
{The median values of the line ratios $^{12}$CO/$^{13}$CO~$J=1-0$ and $^{12}$CO~$J=2-1$/$J=1-0$ are 8.6/6.1 and 0.53/0.39 for the nuclear/disk regions, respectively.
The molecular gas mass derived from $^{13}$CO~$J=1-0$ is strongly correlated with, but systematically lower than, that derived from $^{12}$CO~$J=1-0$.
Most of the galaxies in our sample follow the spatially resolved SF scaling relation between the star formation rate surface density and molecular gas mass surface density, with a median gas depletion timescale of $\sim$1~Gyr. A few galaxies exhibit enhanced SF efficiency, with shorter timescales of $\sim$0.1~Gyr.
Our sample shows a weak correlation between molecular and atomic gas, but a strong correlation between the molecular-to-atomic gas mass ratio ($\rm M_{\rm H_2}/M_{\rm HI}$) and stellar mass, consistent with previous studies.
Galaxies with lower stellar masses in our sample exhibit an excess of atomic gas by one magnitude compared to molecular gas, suggesting that the transformation of atomic gas into molecular gas is less efficient. 
Most galaxies tightly follow the baryonic Tully-Fisher relation, but NGC~2992 and NGC~4594 deviate from the relation due to different physical factors. We find that the ratio of the cold gas (comprising molecular and atomic gas) to the total baryon mass decreases with the gravitational potential of the galaxy, as traced by rotation velocity, which could be due to gas consumption in SF or being heated to the hot phase.
}
{}

\maketitle

\section{Introduction}\label{sec:Introduction}

Molecular gas clouds serve as the cradles of stars, with their distribution being closely related to the star formation rate (SFR). \citet{Kennicutt98} and \citet{Schmidt59} proposed that regions with higher density of total neutral hydrogen gas (H${_2}$ + \ion{H}{I}) correlate with higher SFR. However, studies focusing on the resolved Kennicutt–Schmidt SF law (SFL) reveal that the SFR primarily follows a strong relationship with molecular gas surface density, all existing within a relatively smooth \ion{H}{I} distribution (e.g., \citealt{Leroy08, Bigiel08}).

Molecular and atomic gas content varies across galaxies and is related to stellar mass, morphology, SFR, stellar surface density and color ($NUV-r$ or $g-r$) (e.g., \citealt{Saintonge11, Saintonge22, Tacconi20}). The gas content is usually expressed in terms of the molecular gas mass to stellar mass ($\rm M_{H_2}$/$\rm M_{star}$), atomic gas mass to stellar mass ($\rm M_{HI}$/$\rm M_{star}$), and $\rm M_{H_2}$/$\rm M_{HI}$. Statistical studies based on single-pointing observations of galaxy samples have revealed that the $\rm M_{H_2}$/$\rm M_{HI}$ increases gradually with the $\rm M_{\star}$ (e.g., \citealt{Saintonge17, Boselli14, Bothwell14, Cicone17}). Additionally, the study by \citet{Catinella18} highlighted that changes in $\rm M_{H_2}$/$\rm M_{HI}$ in main-sequence galaxies are primarily driven by variations in the \ion{H}{I} reservoir and also confirmed the significance of different galaxy structures on $\rm M_{H_2}$/$\rm M_{HI}$ changes. The proportion of these cold baryonic components may vary across different galaxies. The multi-wavelength data in our sample provide an opportunity for further investigation.

Our project builds on the CHANG-ES project, which involves systematic VLA observations of 35 nearby edge-on spiral galaxies to investigate the relationship between radio coronas, galactic disks, and galactic environments (e.g., \citealt{Irwin12a, Irwin24}). In addition to high-resolution multi-band VLA observations, these galaxies also have spatially-resolved data across multi wavelengths. For instance, combining new $\rm H\alpha$ images with WISE $22~\mu m$ data provides more reliable estimates of the SFR for the sample (e.g., \citealt{Wiegert15, Vargas19}). The subsequent spatially resolved $\rm H\alpha$ spectral observations are being conducted (e.g., \citet{Lu24, Li24}). High resolution images of \ion{H}{I} are obtained by subtracting the continuum emission from VLA L band observations (e.g., \citealt{Zheng22}), which provide the distribution and proportion of atomic gases. Some well-known galaxies in this sample have already been observed and analyzed the molecular gas in many literatures, such as NGC~660 (e.g., \citealt{Israel09b}), NGC~891 (e.g., \citealt{Yim11}), and NGC~3079 (e.g., \citealt{Irwin92, Israel09a}). The remaining galaxies either lack observations or have only single-pointing observations of their galactic nuclei. To enhance our understanding of the distribution and properties of molecular gas, its correlation with SFR, its contribution to the baryonic gas budget, and its interactions with the galactic halo, it is essential to conduct spatially resolved observations and analyses of molecular gas across this sample.

We conduct the CO-CHANGES project, which simultaneously cover the $^{12}$CO~$J=1-0$, $^{13}$CO~$J=1-0$ and $^{12}$CO~$J=2-1$ emission lines, using the IRAM 30m telescope. The $^{12}$CO~$J=1-0$ line ($\rm T_{ex} \sim 10~K$; $\rm n_{cir} \sim 300~cm^{-3}$) \citep{Evans99} is the best tracer of molecular gas. By combining the three emission lines, we can calculate and analyze the distribution of H$_2$ gas related to star formation and its physical properties, such as temperature and optical depth (e.g., \citealt{Frerking82, Wilson13}), which vary across different SF environments, including the nucleus and the disk. These observations are made at selected positions along the galactic disk of 23 out of the 35 CHANG-ES galaxies, including NGC~4594, which is analyzed as a case study in CO-CHANGES Paper~I \citep{Jiang24}. Table~\ref{table:Parameters} lists the CO-CHANGES sample galaxies along with the relevant parameters from previous studies of the CHANG-ES project (e.g., \citealt{Irwin12a, Wiegert15, Li16a, Vargas19}) which are referenced in this paper. When deriving physical parameters such as the molecular gas temperature and optical depth, we adopt the assumptions and formulas from Paper~I, except where deviations are explicitly discussed.

In this paper, we will introduce the IRAM 30m observation and data reduction in Sect.~\ref{sec:observation}, and present the main results in Sect.~\ref{sec:results}. In Sect.~\ref{sec:discussion}, we will discuss the correlation analysis of molecular gas, atomic gas, and stellar mass within the sample, as well as the SFR and the gas baryon budget. Finally, we will summarize the major conclusions in Sect.~\ref{sec:summary}. We note that 3~$\sigma$ threshold is used to determine whether the spectrum is detected, and the errors drawn in the figure are quoted at 1~$\sigma$ level throughout the paper.

\renewcommand{\arraystretch}{1.5} 
\begin{table*}
\caption{Parameters of the sample galaxies.} 
\footnotesize
\begin{tabular}{ccccccccc}
\hline\hline
Name & Distance & $\rm TC$ & $\rm~V_{rot}$ & Diameter & $\rm M_{\star}$ &  $\rm M_{HI}$  & $\rm SFR$ &  $\rm \Sigma_{SFR}$ \\

& (Mpc) &  & $(\rm km~s^{-1})$ & (kpc) & $(\rm \times10^{10}~M_{\odot})$ & $(\rm \times10^{10}M_{\odot})$  & $(\rm M_{\odot}~yr^{-1})$  & $\rm (\times10^{-3}M_{\odot}~yr^{-1}~kpc^{-2})$ \\
 (1) & (2) & (3) & (4) & (5) & (6) & (7) & (8) & (9)\\
\hline

NGC2613 & $23.4$ & $3.0\pm0.3$ & $290.6$ & $33.7$ & $12.0\pm0.2$ & $0.712^a$  & $3.36\pm0.35$ & $3.77\pm0.39$ \\
NGC2683 & $6.27$ & $3.0\pm0.3$ & $202.6$ & $9.39$ & $1.49\pm0.02$ & $0.052^b$  & $0.25\pm0.03$ & $3.54\pm0.40$ \\
NGC2820 & $26.5$ & $5.0\pm0.5$ & $162.8$ & $13.8$ & $0.467\pm0.013$ & $1.424^c$  & $1.35\pm0.14$ & $9.00\pm0.96$ \\
NGC2992 & $34.0$ & $1.0\pm0.3$ & $58.7$ & $12.5$ & $5.26\pm0.09$ & $0.738^d$  & $5.91\pm0.54$ & $48.2\pm4.4$ \\
NGC3003 & $25.4$ & $4.0\pm0.6$ & $120.6$ & $27.7$ & $0.485\pm0.010$ & $1.072^b$  & $1.56\pm0.16$ & $2.59\pm0.27$ \\
NGC3044 & $20.3$ & $5.0\pm0.5$ & $152.6$ & $18.1$ & $0.660\pm0.013$ & $0.363^b$  & $1.75\pm0.16$ & $6.79\pm0.60$ \\
NGC3432 & $9.42$ & $9.0\pm0.3$ & $109.9$ & $9.96$ & $0.100\pm0.002$ & $0.222^c$  & $0.51\pm0.06$ & $6.36\pm0.69$ \\
NGC3448 & $24.5$ & $5.0\pm0.0$ & $119.5$ & $12.5$ & $0.564\pm0.011$ & $0.832^b$  & $1.78\pm0.18$ & $14.5\pm1.5$ \\
NGC3556 & $14.1$ & $6.0\pm0.3$ & $153.2$ & $24.9$ & $2.81\pm0.04$ & $0.479^b$  & $3.57\pm0.30$ & $7.32\pm0.62$ \\
NGC3735 & $42.0$ & $5.0\pm0.5$ & $241.1$ & $34.4$ & $14.9\pm0.2$ & $1.160^c$  & $6.23\pm0.57$ & $6.71\pm0.61$ \\
NGC3877 & $17.7$ & $5.0\pm0.3$ & $155.1$ & $18.5$ & $2.74\pm0.04$ & $0.148^b$  & $1.35\pm0.12$ & $5.04\pm0.44$ \\
NGC4096 & $10.3$ & $5.0\pm0.3$ & $144.8$ & $11.8$ & $0.613\pm0.010$ & $0.145^b$  & $0.71\pm0.08$ & $6.52\pm0.77$ \\
NGC4157 & $15.6$ & $3.0\pm0.5$ & $188.9$ & $16.6$ & $2.92\pm0.04$ & $0.525^b$  & $1.76\pm0.18$ & $8.15\pm0.83$ \\
NGC4217 & $20.6$ & $3.0\pm0.3$ & $187.6$ & $23.4$ & $4.74\pm0.07$ & $0.275^b$  & $1.89\pm0.18$ & $4.40\pm0.42$ \\
NGC4244 & $4.40$ & $6.0\pm0.3$ & $97.8$ & $14.8$ & $0.090\pm0.002$ & $0.067^c$  & $0.06\pm0.01$ & $0.37\pm0.04$ \\
NGC4302 & $19.4$ & $5.0\pm0.5$ & $167.4$ & $3.82$ & $3.43\pm0.05$ & $0.174^b$  & $0.92\pm0.08$ & $2.52\pm0.21$ \\
NGC4594 & $12.7$ & $1.0\pm0.3$ & $408.0$ & $22.0$ & $26.1\pm0.4$ & $0.033^c$  & $0.43\pm0.04$ & $1.14\pm0.10$ \\
NGC4666 & $27.5$ & $5.0\pm0.3$ & $192.9$ & $32.3$ & $12.5\pm0.2$ & $1.096^b$  & $10.5\pm0.9$ & $12.8\pm1.1$ \\
NGC4845 & $17.0$ & $2.0\pm0.3$ & $176.0$ & $10.4$ & $2.89\pm0.05$ & $0.009^c$  & $0.62\pm0.06$ & $7.33\pm0.68$ \\
NGC5084 & $23.4$ & $-2.0\pm0.3$ & $312.9$ & $4.99$ & $12.3\pm0.2$ & $0.912^b$  & $-$ & $-$ \\
NGC5297 & $40.4$ & $4.5\pm0.5$ & $189.5$ & $25.9$ & $3.69\pm0.07$ & $1.992^e$  & $3.00\pm0.33$ & $5.70\pm0.62$ \\
NGC5792 & $31.7$ & $3.0\pm0.3$ & $208.6$ & $23.7$ & $8.89\pm0.16$ & $1.413^b$  & $4.41\pm0.37$ & $10.0\pm0.8$ \\
UGC10288 & $34.1$ & $5.3\pm0.6$ & $167.1$ & $21.3$ & $2.03\pm0.05$ & $0.955^b$  & $0.66\pm0.07$ & $1.85\pm0.21$ \\

\hline
\end{tabular}\label{table:Parameters}\\

\textbf{Notes.} Galaxy parameters. (1) Galaxy name. (2) Distance is obtained from \citet{Vargas19}. (3) $\rm TC$, the morphological type code from \citet{deVaucouleurs91}, except for NGC~3448, which is from \citet{Huchra12}. (4) $\rm V_{rot}$ is the maximum atomic gas rotation velocity corrected for inclination obtained from HyperLeda (\url{http://leda.univ-lyon1.fr/}), as also noted in \citet{Makarov14}. (5) Diameter is derived from $\rm 22~\mu m$ data by \citet{Wiegert15}. (6) $\rm M_{\star}$ is the stellar mass from \citet{Li16a}, which is derived from the $\emph{2MASS}$ $K$-band apparent magnitude \citep{Skrutskie06}. (7) $\rm M_{HI}$, the total atomic gas mass, is directly available for 14 galaxies from \citet{Zheng22}. The source of the \ion{H}{I} flux is indicated by superscript, and the detailed calculation methods are summarized in Sect.~\ref{subsec:mass2mass}. References and corresponding observational instruments are as follows: a. \citet{Chaves01} with VLA L-band CnB-array; b. \citet{Zheng22} with VLA L-band C-array; c. \citet{Courtois15}, who uniformly processed data from various telescope (NGC~2820 and NGC~3735 from Green Bank 42m; NGC~3432, NGC~4244 and NGC~4594 from Robert C.Byrd Green Bank Telescope; NGC~4845 from Arecibo with line feed system.); d. \citet{Huchtmeier82} with NRAO 91m; e. \citet{Davis83} with NRAO 91m. (8) SFR is the revised star formation rate from \citet{Vargas19}, calculated using a combination of $\rm H\alpha$ and $\rm 22~\mu m$ data. (9) $\rm \Sigma_{SFR}$ is the surface density of SFR and calculated using the $\rm 22~\mu m$ diameter and obtained from \citet{Vargas19}.
\end{table*}

\section{Observations and Data Reduction}\label{sec:observation}

\subsection{IRAM 30m observations}\label{subsec:IRAMobs}

We conducted IRAM 30m CO line observations of 23 CHANG-ES galaxies in a few observation runs from August 2018 to March 2019 through the project 063-18 and 189-18 (PI: Jiang-Tao Li). Details of the observations were summarized in Table~\ref{table:obslog}. The observations were taken with the Eight MIxer Receiver (EMIR), with the E90/E230 combination \citep{Carter12}, which simultaneously covers the $^{12}$CO $J=1-0$, $J=2-1$, and $^{13}$CO $J=1-0$ lines. The Superconductor-Insulator-Superconductor (SIS) receivers were used for observations in position switching (PSW) mode, with a minimum offset of the reference position from the source position of $122^{\prime\prime}$. We used the 24 Fourier Transform Spectrometer (FTS) units as backend, covering a total of $32\rm~GHz$ bandwidth across all channels with a frequency resolution of $200\rm~kHz$. The full width at half-maximum (FWHM) of the primary beam of IRAM 30m was $\approx21.4^{\prime\prime}$ and $\approx10.7^{\prime\prime}$ at 115~GHz and 230~GHz, while the main beam efficiency was set to 78 percent and 59 percent at these frequencies, respectively. \footnote{\url{http://www.iram.es/IRAMES/mainWiki/Iram30mEfficiencies}}

We observed 3 to 13 separate positions typically along the disk of each galaxy, in order to study the spatial distribution of the gas. Since most galaxies in our sample are axisymmetric, we can position the beams asymmetrically around the galaxy center, increasing the number of sampling points on one side, which facilitates subsequent kinematic analyses, such as rotation curve fitting. One of these observations in each galaxy is pointed at the galaxy nucleus. Refer to the appendix for the specific distribution of the observation positions. The on-source integration time for most of these positions was 9.8~minutes, while for outer regions of disk we doubled the on-source integration time to increase the signal-to-noise ratio (S/N). The total integration time for all galaxies amounted to 56 hours.

\begin{table*}
\caption{IRAM 30m observation log.} 
\footnotesize
\begin{tabular}{cccccccccccc}
\hline\hline
Galaxy & RA & DEC & Date & $\tau_{225\text{GHz}}$ & $\rm n_{obs}$ & $t_{\text{exp}}$ & $t_{\text{exp}}$ & $t_{\text{exp}}$ & rms & rms & rms \\

& (J2000) & (J2000) & & & & (min) & (min) & (min) & (mk) & (mk) & (mk) \\
 (1) & (2) & (3) & (4) & (5) & (6) & (7) & (8) & (9) & (10) & (11) & (12)\\
\hline
NGC2613-1 & 8:33:18.39 & -22:57:56.73 & 2/25/2019 & 0.26 & 1 & 19.63 & 19.63 & 19.63 & 15.48 & 6.71 & 16.21 \\
NGC2613-2 & 8:33:21.30 & -22:58:15.54 & 2/25/2019 & 0.26 & 1 & 19.63 & 19.63 & 19.63 & 14.94 & 4.99 & 10.71 \\
NGC2613-3 & 8:33:22.75 & -22:58:24.95 & 2/25/2019 & 0.26 & 1 & 19.63 & 19.63 & 19.63 & 14.54 & 5.05 & 10.92 \\
...&&& ...&&&...&&&...&&\\

\hline
\end{tabular}\label{table:obslog}\\

\textbf{Notes.} Listed items (The full version is only available online.): (1) Galaxy name with observation positions. Bold names indicate companion galaxy or non-disk position (possibly halfway to the companion galaxy), with companion galaxies indicated in parentheses. (2) Position RA. (3) Position DEC. (4) Observation date. (5) $\tau_{\rm~225GHz}$ is the average opacity at 225~GHz during observation dates. (6) $\rm n_{obs}$ is the total number of observation scans at each position, with each scan having an effective on-source integration time of 9.8~minutes. (7)-(9) The total on-source integration time at each position for ${\rm ^{12}CO}$ $J=1-0$, ${\rm ^{13}CO}$ $J=1-0$, and ${\rm ^{12}CO}$ $J=2-1$ lines, respectively, including both vertical and horizontal polarization components. (10)-(12) The root mean square (rms), calculated from fitting the baseline of the combined spectra for both polarization components for ${\rm ^{12}CO}$ $J=1-0$, ${\rm ^{13}CO}$ $J=1-0$, and ${\rm ^{12}CO}$ $J=2-1$ lines, respectively.

\end{table*}

\subsection{Data reduction}\label{subsec:DataReduction}

We utilized the CLASS (Continuum and Line Analysis Single-dish software) from the GILDAS package (version jul21) \footnote{\url{http://www.iram.fr/IRAMFR/GILDAS}} to reduce and analyze the IRAM 30m/EMIR data. In CO-CHANGES Paper~I, we used an older version (jan17a) for NGC~4594, which did not result in significant differences.

Data taken with the FTS backend sometimes showed intermittent ``platforming'' issues between individual units \footnote{\url{https://www.iram.fr/~gildas/demos/class/class-tutorial.pdf}}. Following the methodology outlined in CO-CHANGES Paper~I (see also \citet{Lisenfeld19}), we corrected the ``platforming'' issues and binned the $^{12}$CO $J=1-0$, $J=2-1$, and $^{13}$CO $J=1-0$ spectra to a resolution of $\Delta v=10\rm~km~s^{-1}$ for all galaxies. Different from the Paper I, we primarily fitted spectra using a polynomial function of up to the second degree. To combine the spectra from different polarization components and multiple on-source integration times, we used sigma-weighted averaging.

Approximately 22\% of the $^{12}$CO $J=2-1$ spectra were affected by standing waves, mainly from horizontal polarization. For spectra with S/N \textgreater $7\sigma$, we retained the data without correction. For spectra with $3\sigma$ \textless S/N \textless $7\sigma$, we subtracted a sinusoidal function based on conditions from other polarization or observations. This correction, applied to only 1\% of the data, was made when it did not significantly affect the signal and indeed improved the rms. We excluded spectra with S/N \textless $3\sigma$ affected by standing waves, accounting for 11\% of the total $^{12}$CO $J=2-1$ spectra. The rms thresholds varied for each galaxy, with an average of 30~mK. This exclusion typically reduced the rms from approximately 40~mK to 10~mK.

For $^{12}$CO $J=1-0$ and $^{13}$CO $J=1-0$ spectra, all data were retained except for NGC~3735, in which the $^{13}$CO $J=1-0$ spectra at positions ``1'' to ``4'' were not covered due to the incorrect frequency settings during observations on August 24, 2018.

\begin{figure*}
\begin{center}
\includegraphics[width=0.95\textwidth]{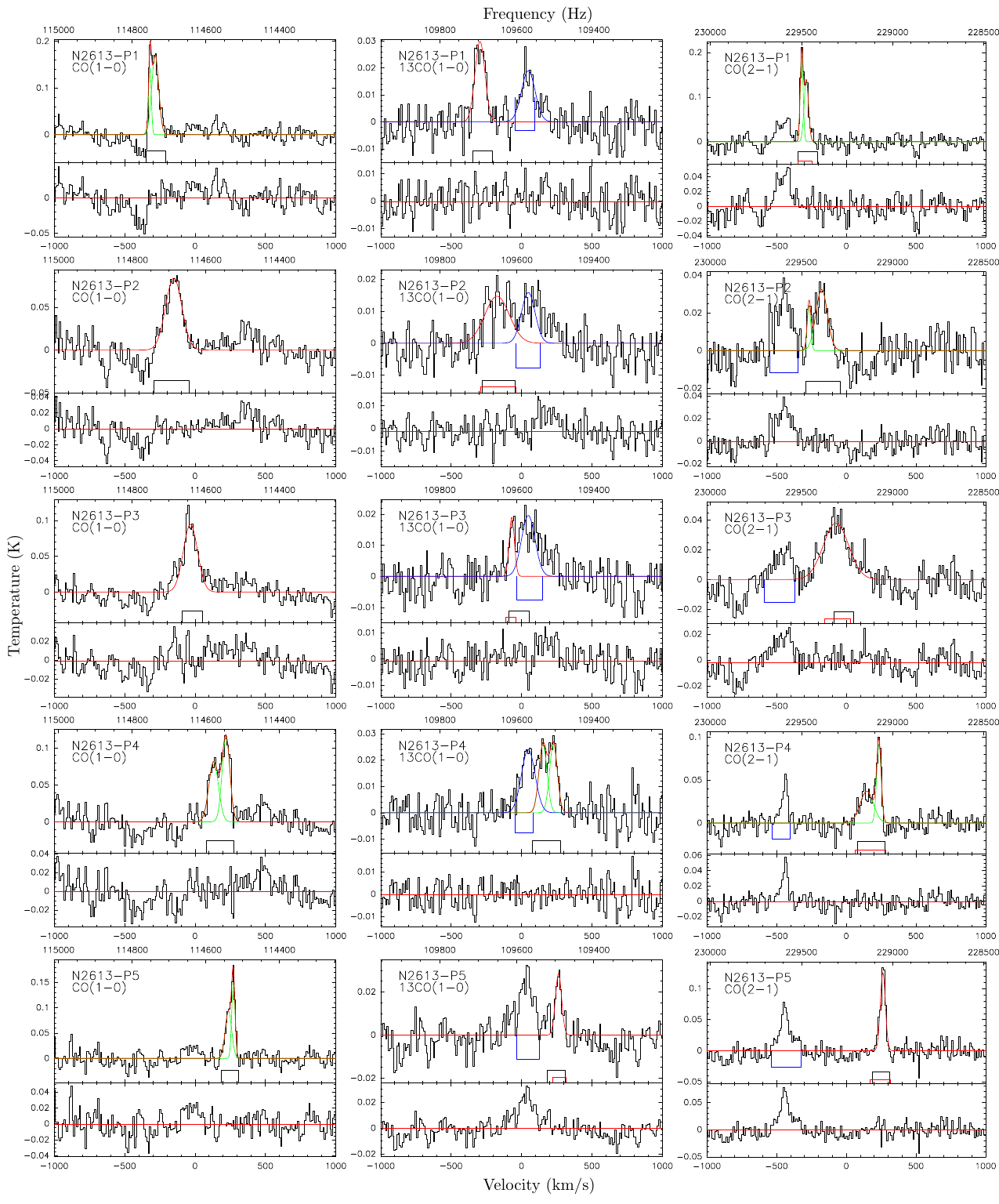}

\caption{$^{12}$CO~$J=1-0$ (left), $^{13}$CO~$J=1-0$ and (middle), and $^{12}$CO~$J=2-1$ (right) spectra of NGC~2613. Each row represents a different locations, as shown in Fig.~\ref{fig:example2613}. In the upper half of each panel, the $y$-axis is the main beam temperature after correcting for the main beam forward efficiency. The lower half of each panel shows the residual of the best-fit. The lower $x$-axis for all galaxies shows the same velocity range within $\pm1000\rm~km~s^{-1}$. The top $x$-axis indicates the frequency range for the three different molecular lines of each galaxy, set according to the zero velocity, which is defined by the galaxy's systematic velocity \citep{Irwin12a}. All spectra are binned to a velocity resolution of $10\rm~km~s^{-1}$. The green curves represent the best-fit Gaussian lines, and the red line is the baseline. The blue curve represents the fitted RFI, as detailed in Sect.~\ref{subsec:DataReduction}. The black box indicates the velocity range ($\Delta W$) determined by the $^{12}$CO~$J=1-0$ linewidth, which is used as the fitting range. This range is also displayed on the $^{13}$CO~$J=1-0$ and $^{12}$CO~$J=2-1$ spectra to show the position of the $^{12}$CO emission line. If there is a deviation, the red box highlights the adjusted velocity range used for fitting the other lines. The blue boxes mask some low-significance features which could be artificial and may slightly affect the fitting. These features are masked from the fitting and the calculation of the rms. 
}\label{fig:2613spec}
\end{center}
\end{figure*}

Unexpected spike features were detected in some spectra, which could affect the analysis of the emission lines. For example, in the $^{13}$CO~$J=1-0$ and $^{12}$CO~$J=2-1$ spectra of NGC~2613 and NGC~3044, we observed abnormal spikes at fixed rest frequencies that do not shift with redshift, as shown in Fig.~\ref{fig:2613spec} within the velocity range indicated by the blue box. In NGC~3044, these spikes were only detected on specific dates. We therefore attributed these signals to RFI (Radio Frequency Interference) and subsequently masked these spikes. However, for NGC~2613 at positions ``2'', ``3'', and ``4'', the spikes were mixed with the galactic $^{13}$CO~$J=1-0$ lines and could not be easily masked. Hence, we applied a Gaussian fit (the blue curve in Fig.~\ref{fig:2613spec}) to the spike at position ``1'', fixing its line width and intensity, to subtract the spikes from the other three positions (``2'', ``3'', and ``4''). We applied 3$\sigma$ to these three positions and excluded them from the statistical analysis. The fitting results were presented in Table~\ref{table:COparameter}. For the remaining galaxies, we did not correct for RFI as it did no affect on the lines of interest.

\begin{figure*}
\begin{center}
\includegraphics[width=0.95\textwidth]{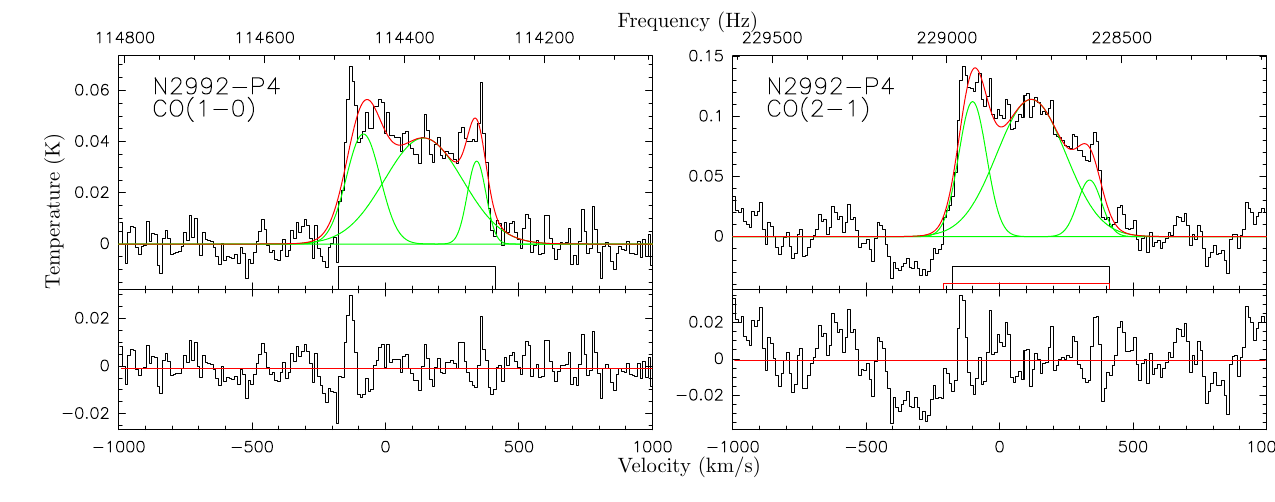}
\caption{The spectra of $^{12}$CO~$J=1-0$ (left panel) and $^{12}$CO~$J=2-1$ (right panel) for NGC~2992 are presented in the same format as in Fig.~\ref{fig:example2613}.
}\label{fig:2992nucleus}
\end{center}
\end{figure*}

In our sample, five galaxies exhibited $^{12}$CO~$J=1-0$ emission in their nuclei with a total velocity width exceeding 400$\rm~km~s^{-1}$, including NGC~2683, NGC~2992, NGC~3735, NGC~4845, and NGC~5792. In the nucleus of NGC~2992, the spectra reveal a double-peaked structure with a total emission line width of 587$\rm~km~s^{-1}$ for $^{12}$CO $J=1-0$ and 620$\rm~km~s^{-1}$ for $^{12}$CO $J=2-1$ (see Fig.~\ref{fig:2992nucleus}), primarily driven by the rotation of the disk in the nuclear region. There is an extremely active AGN at the center of NGC~2992 (\citealt{Gilli00}), with multi-wavelength evidence revealing AGN-driven outflows and winds (e.g., \citealt{Irwin17, Zanchettin23}).

We fitted the CO emission lines with multiple (up to three) Gaussian components to determine the centroid velocities and calculated the integrated intensities of different CO lines for each galaxy by summing the area over the velocity range of $\Delta W$. As $^{12}$CO~$J=1-0$ is the strongest line in most cases, its line emission velocity range is marked by the black box in Fig.~\ref{fig:2613spec}. For the $^{13}$CO$J=1-0$ and $^{12}$CO~$J=2-1$ lines, which are not detected, we adopt the same velocity range as $^{12}$CO~$J=1-0$. However, when these lines are detected with a S/N greater than 3$\sigma$, the velocity window is determined based on their actual line shapes and shown as the red box. Subsequently, we calculated the ratios between different CO emission lines, considering the beam dilution factor to correct for intensity variations caused by differences in the half power beam width at different frequencies, consistent with the method used in Paper I. The beam dilution factors for the $^{12}$CO $J=1-0/^{13}$CO $J=1-0$ and $^{12}$CO $J=1-0/^{12}$CO $J=2-1$ ratios typically ranged from 0.87 to 1.08 (median 0.92) and from 0.34 to 0.61 (median 0.42), respectively. An exception was made for the $^{12}$CO $J=1-0/^{12}$CO $J=2-1$ ratio in NGC~4845 and NGC~5792, where the beam dilution correction factor ranged from 0.37 to 1.15 and 0.32 to 3.15, respectively, due to the exceptionally strong CO emission in both lines from the galactic nucleus.

When we estimated the optical depth ($\tau_{\rm ^{13}CO}$) and kinetic temperature ($\rm T_{K}$) of the molecular gas under LTE (the local thermal equilibrium) conditions, we adopted the following assumptions. We assumed that both $^{12}$CO~$J=1-0$ and $^{13}$CO~$J=1-0$ have the same excitation temperature and filling factor, with $\tau_{\rm ^{12}CO} \gg 1$ and $\tau_{\rm ^{13}CO} \ll 1$. These assumptions were generally valid for galaxies in our sample that do not exhibit extreme conditions such as intense nuclear starbursts. With the abundance ratio of $^{12}$CO/$^{13}$CO generally ranging from 60 to 90 in molecular clouds \citep{Hollenbach87}, we followed \citet{Cormier18} by adopting a value of 60. When calculating the kinetic temperature, we assumed an abundance ratio of $^{12}$CO/H$_2$ of $8 \times 10^{-5}$ \citep{Frerking82}. Detailed formulas for these calculations are provided in Paper~I, and the results are summarized in Table~\ref{table:COparameter}. The calculation of uncertainties follows the method described in Paper I, employing 5000 boot-strap samples to determine the distributions of each measurement. All uncertainties are estimated statistically from the measurements.

\section{Result}\label{sec:results}

\subsection{Radial distribution and total mass of molecular gas}\label{subsec:Intensityandmass}

In this subsection, we present the spatial distribution of the $^{12}$CO $J=1-0$, $J=2-1$, and $^{13}$CO $J=1-0$ lines along the galactic disk, and compare the total molecular gas mass ($\rm M_{H_2}$) estimated from the $^{12}$CO and $^{13}$CO $J=1-0$ lines. The radial CO line and H$_2$ column density profiles for each galaxy are provided in the appendix, with an example of NGC~2613 shown in Fig.~\ref{fig:example2613}. For almost galaxies, the line intensity decreases with radius.

\begin{figure*}
\begin{center}
\includegraphics[width=0.55\textwidth]{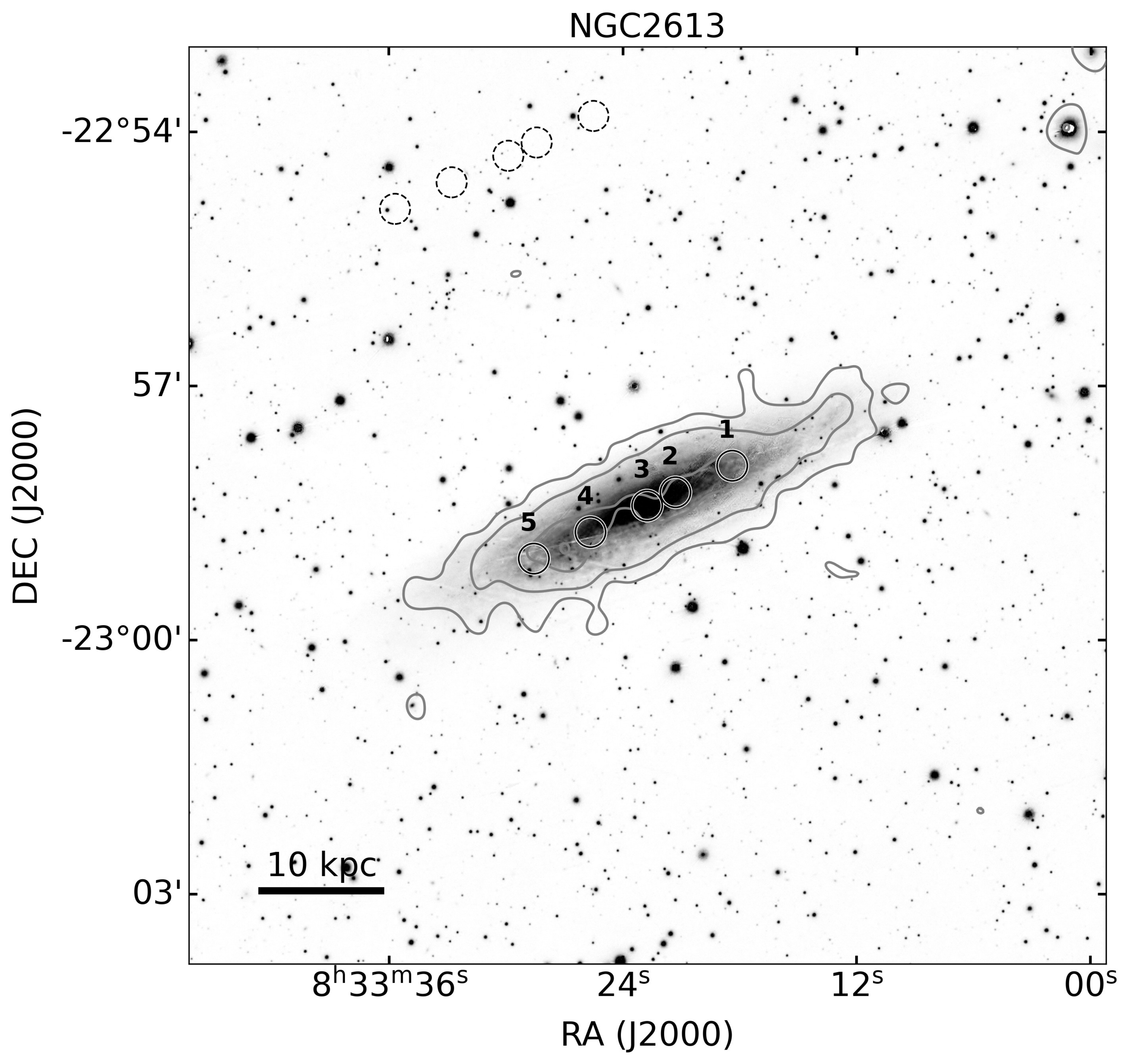}
\hfill
\includegraphics[width=0.40\textwidth]{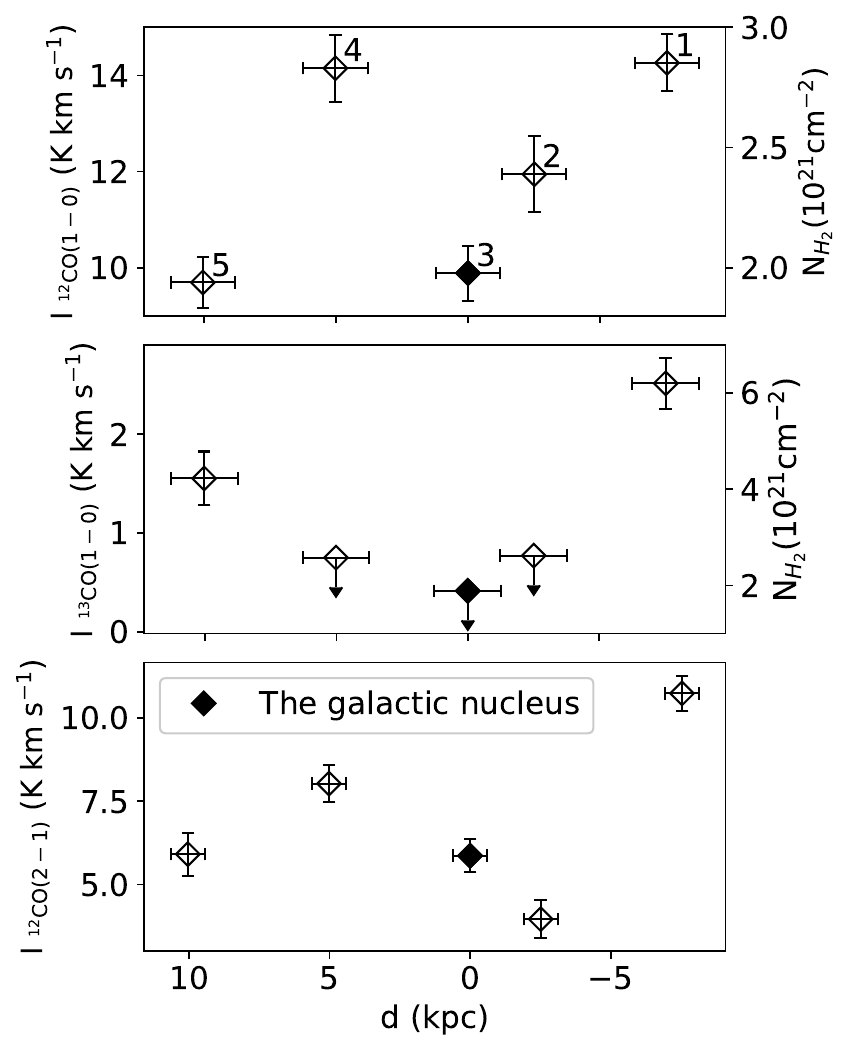}
\caption{
$Left~panel$: The $Pan-STARSS$ $g$-band image displays an 11 arcmin $\times$ 11 arcmin area centered on NGC~2613. The solid circles indicate the location of IRAM 30m beams with a $\approx21.4^{\prime\prime}$ diameter for the $^{12}$CO $J=1-0$ spectral band. The dashed circles represent the background location for the PSW observations. The overlay contours are from a convolved $\rm 22~\mu m$ image from the WISE archive. For most galaxies, contour levels include 3$\sigma$, 10$\sigma$, and 50$\sigma$, with additional levels at 100$\sigma$ and 150$\sigma$ for those with more pronounced flux peaks, where 3$\sigma$ corresponds to a median flux of 0.01~mJy. UGC~10288 is an exception, where the contour levels are $3\sigma$ and $10\sigma$. $Right~panel$: The right panel displays the integrated intensities of the $^{12}$CO $J=1-0$ (top row), $^{13}$CO $J=1-0$ (middle row) and $^{12}$CO $J=2-1$ (bottom row) lines along the galaxy disk, corresponding to the numbered positions in the left panel. The filled diamond indicates the galaxy nucleus. The x-axis represents the projected distance to the minor axis as listed in Table~\ref{table:COparameter}. The right y-axes of the top two panels show the molecular gas column density $N_{\rm H_2}$, derived from the corresponding line.
}\label{fig:example2613}
\end{center}
\end{figure*}

\begin{table*}
\caption{Observed and derived parameters of the CO lines.} 
\footnotesize
\begin{tabular}{
    cccccccccccc
}

\hline\hline
Galaxy & $d$ & $\rm~I_{\rm ^{12}CO_{\rm 10}}$ & $v_{\rm ^{12}CO_{\rm 10}}$ & $\rm~I_{\rm ^{13}CO_{\rm 10}}$ &$v_{\rm ^{13}CO_{\rm 10}}$& $\rm~I_{\rm ^{12}CO_{\rm 21}}$ &$v_{\rm ^{12}CO_{\rm 21}}$& ${\rm \mathsmaller{\mathsmaller{\frac{^{12}CO_{\rm 10}}{^{13}CO_{\rm 10}}}}}$ & ${\rm \mathsmaller{\mathsmaller{\frac{^{12}CO_{\rm 21}}{^{12}CO_{\rm 10}}}}}$ & $\tau(\rm ^{13}CO)$ &  $T_{\rm K}$\\

 & (kpc) & ($\rm K~km~s^{-1}$) &($\rm km~s^{-1}$) & ($\rm K~km~s^{-1}$) &($\rm km~s^{-1}$)& ($\rm~K~km~s^{-1}$ )&($\rm km~s^{-1}$)&  &  & & (K) \\
 (1) & (2) & (3) & (4) & (5) & (6) & (7) & (8) & (9) & (10) & (11) & (12)\\
\hline
N2613-1 & -7.5 & $14.3\pm0.6$ & $-291.5^{+2.6}_{-2.4}$ & $2.5\pm0.3$ & $-294.8\pm4.6$ & $10.7\pm0.5$ & $-302.6^{+3.9}_{-3.0}$ & $5.2^{+0.3}_{-0.5}$ & $0.28\pm0.01$ & $0.21^{+0.02}_{-0.06}$ & $19.4^{+1.4}_{-2.3}$  \\
N2613-2 & -2.5 & $11.9\pm0.8$ & $-150.6\pm4.5$ & $\textless0.8$ & $-173.4\pm8.6$ & $4.0\pm0.6$ & $-193.4^{+7.6}_{-6.4}$ & $ - $ & $0.15^{+0.01}_{-0.03}$ & $ - $ & $ - $  \\
N2613-3 & 0.0 & $9.9\pm0.6$ & $-32.0\pm3.7$ & $\textless0.6$ & $-68.6\pm5.0$ & $5.9\pm0.5$ & $-73.9\pm8.2$ & $ - $ & $0.24^{+0.01}_{-0.02}$ & $ - $ & $ - $  \\
...&&& ...&&&...&&&...&&\\

\hline
\end{tabular}\label{table:COparameter}\\

\textbf{Notes.} Listed items (The full version is only available online.): (1) Galaxy labels for different observational regions, as shown in Fig.~\ref{fig:example2613}. The bold name indicate companion galaxies or locations off the main disk, as listed in Table~\ref{table:obslog}. (2) $d$ denotes the projected distance to the galaxy's minor axis. (3), (5), (7) The integrated line intensities for $^{12}$CO~$J=1-0$, $^{13}$CO~$J=1-0$ and $^{12}$CO~$J=2-1$ lines, corrected for main beam and forward efficiencies. If the lines are not firmly detected, we list the $3\sigma$ upper limit, and use '-' in the velocity column and other parameter calculations. (4), (6), (8) The centroid velocities are derived from the intensity-weighted average velocity of the Gaussian components. (9), (10) The intensity line ratios of $^{12}$CO~$J=1-0$ to $^{13}$CO~$J=1-0$ and $^{12}$CO~$J=2-1$ to $^{12}$CO~$J=1-0$ are corrected for beam dilution as discussed in Sect.~\ref{subsec:DataReduction}. (11), (12) $\tau(\rm ^{13}CO)$ and $\rm T_{K}$ represent the optical depth and kinetic temperature, respectively, under local thermodynamic equilibrium (LTE). The uncertainties only account for the statistical errors from the measurements.

\end{table*}

Both $^{12}$CO $J=1-0$ and $^{13}$CO $J=1-0$ are used to measure the molecular gas column density, with the former usually stronger but optically thick.
$^{13}$CO $J=1-0$ is typically regarded as optically thin and a reliable tracer of molecular hydrogen in regions with a higher proportion of dense gas. However, in the normal galactic disk, it is less sensitive to the diffuse phase of molecular gas (\cite{Cormier18}). \cite{Cormier18} conducted a study on the spatially resolved $^{13}$CO(1-0)-to-H$_2$ conversion factor in disks of nearby galaxies, finding that the average value for the entire galaxy sample is $X_{\rm^{13}CO(1-0)}=1.0\times10^{21}\rm~cm^{-2}/(K~km~s^{-1})$, with an uncertainty of approximately a factor of 2 due to calibration uncertainties, including variations in CO line ratios, dust properties, and stellar populations.

We employ an $X$ factor of $X_{\rm ^{12}CO(1-0)}=2\times10^{20}\ \rm cm^{-2}/(K\ km\ s^{-1})$ for $^{12}$CO $J=1-0$ to calculate the molecular gas column density $N_{\rm H_2}$. The $X$ factor generally varies with several parameters of a molecular cloud, such as temperature, optical depth, column density, and metallicity. In general, there is good agreement on the estimated $X$ factors through various independent predictions and observations, such as estimates from the dynamics of molecular clouds (e.g., \citealt{Solomon87}), isotopologues (e.g., \citealt{Goldsmith08}), extinction (e.g., \citealt{Frerking82}), dust emission (e.g., \citealt{Dame01}), and gamma-ray emission (e.g., \citealt{Strong96}). These methods yield an $X$ factor of approximately $X_{\rm ^{12}CO(1-0)} \approx (1\sim4) \times 10^{20}\ \rm cm^{-2}/(K\ km\ s^{-1})$ (\citealt{Shetty11b}). No matter these studies focus on larger masses, higher resolution scales, or nearby normal star-forming galaxies, the conclusions are generally consistent with those in the Milky Way, with deviations around 40 percent in extragalactic environments (e.g., \citealt{Rosolowsky06, Bolatto13}). Previous studies have generally concluded that galaxy centers exhibit lower $X$ factors (e.g., \citealt{Sandstrom13,Bolatto13}), likely due to higher temperatures, increased velocity dispersion, and elevated metallicity (e.g.,\citealt{Henry99}). Those effects may become even more pronounced in galaxies with AGN activity at their centers (e.g., \citealt{Papadopoulos99,Yao03}). To maintain consistency with the methodology used for deriving molecular gas masses from $^{13}$CO $J=1-0$, we adopt the same $X$ factor value for both the centers and disks. In our subsequent discussions, we will compare molecular gas mass estimates based on both $^{12}$CO $J=1-0$ and $^{13}$CO $J=1-0$.

We measure the CO line flux at a few isolated positions along the disk of each galaxy. In order to estimate the total molecular gas mass of the galaxy, we need to apply a reasonable scaling factor between our directly measured CO line flux and the same line flux integrated over the whole galaxy. Here we simply assume the CO line flux has a constant ratio to the IR flux, which traces the dust-obscured emissions mainly from SF regions (e.g. \citealt{Kennicutt98,Kennicutt12,Leroy08}). We adopt the $\rm 22~\mu m$ flux from the Wide-field Infrared Survey Explorer (WISE) survey, as it is uniformly available for all our sample galaxies \citep{Vargas19} and is little impacted by the extinction effect (however, see \citealt{Li16a} for the potential extinction effect for edge-on galaxies. In the study by \citet{Vargas18}, edge-on galaxies were found to exhibit an average reduction in $\rm 25~\mu m$ flux by a factor of 1.36 due to extinction when compared to face-on galaxies. This effect has not been accounted for in our analysis.). We reprocess the WISE 22~$\mu$m images by subtracting a local background and convolving them to a resolution of $21^{\prime\prime}$ (the IRAM 30m main beam size at $^{12}$CO $J=1-0$) using convolution kernels provided by \citet{Aniano11}. We calculate the total $\rm 22~\mu m$ flux within the contour defined at a level of three times the background standard deviation ($3\sigma$), which is often comparable in radial extension to the regions covered by our CO line observations. The scaling factor used to correct for the directly measured CO line flux ($\textit{f}_{\text{IR}}$; see Table~\ref{table:masstable}) is defined as the ratio between the $\rm 22~\mu m$ fluxes measured within the IRAM 30m beams (characterized with isolated circular regions) and the $3~\sigma$ IR contour of the galaxy. Here we exclude the nuclear region of the galaxy, as it sometimes shows unusually strong IR emissions from the active galactic nuclei (AGN) that do not correlate to the CO line emissions. The molecular gas mass measured from the beam covering this nuclear region is directly added to the total mass of the galaxy.

The total molecular gas mass, derived from the corrected molecular gas mass in the disk ($\textit{f}_{\text{IR}}$ $\times$ $\rm M_{H_2,disk}^o$) plus the original molecular gas mass from the nuclear region ($\rm M_{H_2,nucleus}^o$), is summarized in Table~\ref{table:masstable}. For NGC~2683, NGC~4244, and NGC~2820, we define a single elliptical region to calculate the correction factor $\textit{f}_{\text{IR}}$, ensuring that only the IR flux from the galaxy itself is included. Specifically, for NGC~2683 and NGC~4244, the 3$\sigma$ IR contours are fragmented into multiple discrete regions across their disks. For NGC~2820, the elliptical region is used to distinguish the galaxy's IR flux from that of its companion galaxies. For consistency, we recalculated the total molecular gas mass of NGC~4594 from Paper~I using the same method, and found negligible difference.

The infrared and radio emissions in an AGN core, where the $\rm 22~\mu m$ flux is heavily influenced by AGN activity, differ from those in star formation regions, so the calculation for the nucleus is flawed. Thus, we apply a modified method to galaxies where the nucleus flux exceeds ten times the average flux of the galaxy disk, including NGC~2992, NGC~3448, NGC~4845, and NGC~5792. For NGC~2992, NGC~3448, and NGC~4845, the total masses of molecular gas are obtained by summing the original molecular gas mass from both the disk and nucleus. However, given the compactness of NGC~5792 and the limitations of IR resolution, we use the H$\alpha$ image from \citet{Vargas19} to derive the scaling factor. The higher resolution H$\alpha$ image helps us to separate the inner ring or other stellar/gaseous structures from the galactic nucleus. Similarly, we then calculate the scaling factor as the ratio of the total H$\alpha$ flux within the IRAM 30m beam or within the $\rm 22~\mu m$ $3\sigma$ contour, both excluding the flux from the nuclear region. This approach ensures consistency with the method applied to other galaxies in terms of area definition, while the scaling factor itself is based entirely on the H$\alpha$ flux.

\renewcommand{\arraystretch}{1.5} 
\begin{table*}
\caption{Molecular gas mass of the sample galaxies.} 
\footnotesize
\begin{tabular}{cccccccc}
\hline\hline
Name & $\textit{f}_{\text{IR}}$ & $\rm M_{H_2,disk}^o$ & $\rm M_{H_2,disk}$ & $\rm M_{H_2,nucleus}^o$ & $\rm M_{H_2}$ &  $\rm M_{H_2,^{13}CO}$  &  $\rm \Sigma_{M_{H_2}} $ \\

&  & $(\rm \times10^{8}M_{\odot})$ & $(\rm \times10^{8}M_{\odot})$ & $(\rm \times10^{8}M_{\odot})$ & $(\rm \times10^{8}M_{\odot})$ & $(\rm \times10^{8}M_{\odot})$   & $\rm (M_{\odot}~pc^{-2})$ \\
 (1) & (2) & (3) & (4) & (5) & (6) & (7) & (8) \\
\hline

NGC2613 & $6.79$ & $7.47^{+0.88}_{-0.84}$ & $50.7^{+6.0}_{-5.7}$ & $1.48\pm0.33$ & $52.3\pm5.6$ & $33.2^{+9.4}_{-8.7}$ & $5.86^{+0.67}_{-0.56}$  \\
NGC2683 & $5.53$ & $0.31^{+0.03}_{-0.02}$ & $1.73^{+0.15}_{-0.13}$ & $0.26^{+0.03}_{-0.04}$ & $1.99^{+0.15}_{-0.13}$ & $1.69^{+0.36}_{-0.32}$ & $2.87^{+0.22}_{-0.20}$  \\
NGC2820 & $4.09$ & $1.21^{+0.31}_{-0.32}$ & $4.95^{+1.28}_{-1.31}$ & $1.73^{+0.27}_{-0.28}$ & $6.70^{+1.14}_{-1.09}$ & $5.60^{+1.79}_{-1.47}$ & $4.53^{+0.81}_{-0.68}$  \\
NGC2992 & $-$ & $1.47^{+0.88}_{-0.53}$ & $-$ & $7.55^{+1.11}_{-1.06}$ & $9.08^{+1.32}_{-1.36}{}^\star$ & $6.00^{+3.02}_{-1.88}{}^\star$ & $7.33^{+0.93}_{-1.23}$  \\
NGC3003 & $4.88$ & $0.90^{+0.19}_{-0.16}$ & $4.42^{+0.94}_{-0.78}$ & $0.94^{+0.29}_{-0.33}$ & $5.36^{+1.08}_{-0.88}$ & $8.79^{+2.89}_{-2.36}$ & $0.89^{+0.18}_{-0.14}$  \\
NGC3044 & $3.60$ & $1.13^{+0.14}_{-0.16}$ & $4.07^{+0.52}_{-0.59}$ & $1.26^{+0.18}_{-0.20}$ & $5.32^{+0.55}_{-0.61}$ & $4.54^{+1.42}_{-1.02}$ & $2.06^{+0.21}_{-0.25}$  \\
NGC3432 & $4.42$ & $0.15^{+0.04}_{-0.03}$ & $0.67^{+0.16}_{-0.15}$ & $0.06\pm0.03$ & $0.73^{+0.17}_{-0.15}$ & $1.10^{+0.42}_{-0.32}$ & $0.95^{+0.22}_{-0.19}$  \\
NGC3448 & $-$ & $0.83^{+0.26}_{-0.21}$ & $-$ & $0.71^{+0.32}_{-0.36}$ & $1.56^{+0.46}_{-0.44}{}^\star$ & $\textless2.3{}^\star$ & $1.28^{+0.39}_{-0.36}$  \\
NGC3556 & $4.71$ & $6.31\pm0.15$ & $29.7\pm0.7$ & $1.90\pm0.08$ & $31.6^{+0.7}_{-0.6}$ & $14.1\pm1.7$ & $6.48^{+0.13}_{-0.12}$  \\
NGC3735 & $3.34$ & $16.9^{+1.3}_{-1.2}$ & $56.6^{+4.3}_{-3.9}$ & $14.9^{+1.1}_{-1.3}$ & $71.5^{+3.9}_{-4.0}$ & $42.2^{+8.9}_{-8.1}$ & $7.67^{+0.39}_{-0.41}$  \\
NGC3877 & $7.54$ & $2.66^{+0.13}_{-0.15}$ & $20.0^{+1.0}_{-1.1}$ & $2.07^{+0.16}_{-0.14}$ & $22.1^{+1.0}_{-1.1}$ & $14.3^{+2.4}_{-2.6}$ & $8.29^{+0.37}_{-0.44}$  \\
NGC4096 & $7.60$ & $0.61\pm0.04$ & $4.60^{+0.28}_{-0.31}$ & $0.51\pm0.04$ & $5.11^{+0.31}_{-0.32}$ & $2.66^{+0.61}_{-0.60}$ & $4.70^{+0.27}_{-0.28}$  \\
NGC4157 & $3.80$ & $3.65^{+0.21}_{-0.22}$ & $13.9\pm0.8$ & $1.27\pm0.15$ & $15.1\pm0.8$ & $10.2\pm1.7$ & $6.95^{+0.28}_{-0.40}$  \\
NGC4217 & $3.40$ & $10.8\pm0.2$ & $36.8\pm0.8$ & $4.59^{+0.23}_{-0.25}$ & $41.4\pm0.8$ & $28.1^{+2.2}_{-2.4}$ & $9.66^{+0.17}_{-0.19}$  \\
NGC4244 & $3.53$ & $0.013\pm0.003$ & $0.05\pm0.01$ & $0.003\pm0.001$ & $0.05\pm0.01$ & $0.14^{+0.04}_{-0.03}$ & $0.029\pm0.006$  \\
NGC4302 & $4.85$ & $3.92^{+0.17}_{-0.21}$ & $19.0^{+0.8}_{-1.0}$ & $2.55\pm0.20$ & $21.6\pm0.9$ & $13.8\pm1.8$ & $5.90^{+0.22}_{-0.28}$  \\
NGC4594 & $4.66$ & $1.00^{+0.08}_{-0.09}$ & $4.66^{+0.37}_{-0.42}$ & $0.05\pm0.01$ & $4.71^{+0.35}_{-0.39}$ & $6.51^{+1.30}_{-1.12}$ & $1.23^{+0.08}_{-0.10}$  \\
NGC4666 & $5.83$ & $18.2\pm0.4$ & $105.9^{+2.2}_{-2.1}$ & $14.9^{+0.4}_{-0.5}$ & $120.8^{+2.2}_{-2.1}$ & $59.5^{+5.7}_{-5.0}$ & $14.8\pm0.3$  \\
NGC4845 & $-$ & $1.13^{+0.15}_{-0.16}$ & $-$ & $4.90^{+0.11}_{-0.15}$ & $6.03\pm0.18{}^\star$ & $3.86^{+0.57}_{-0.56}{}^\star$ & $7.13\pm0.22$  \\
NGC5084 & $-$ & $-$ & $-$ & $-$ & $\textless~1.3$ & $\textless~3.9$ & $-$  \\
NGC5297 & $3.38$ & $5.24^{+0.52}_{-0.54}$ & $17.7\pm1.8$ & $2.17^{+0.42}_{-0.45}$ & $19.9^{+1.7}_{-1.8}$ & $14.6^{+4.7}_{-4.4}$ & $3.80^{+0.28}_{-0.33}$  \\
NGC5792 & $2.83{}^{\star\star}$ & $11.3^{+0.8}_{-0.9}$ & $32.1^{+2.3}_{-2.6}$ & $11.8\pm0.7$ & $44.0^{+2.5}_{-2.7}{}^{\star}{}^{\star}$ & $29.1^{+5.5}_{-4.9}{}^{\star}{}^{\star}$ & $9.97^{+0.52}_{-0.58}$  \\
UGC10288 & $2.46$ & $4.55^{+0.39}_{-0.53}$ & $11.2^{+1.0}_{-1.3}$ & $1.96^{+0.42}_{-0.47}$ & $13.2^{+1.0}_{-1.1}$ & $11.0^{+3.4}_{-2.4}$ & $3.69^{+0.32}_{-0.37}$  \\

\hline
\end{tabular}\label{table:masstable}\\

\textbf{Notes.} Galaxies parameters. (1) Galaxy name. (2) $\textit{f}_{\text{IR}}$ is the scaling factor derived from the $\rm 22~\mu m$ flux within $\rm 3\sigma$ contour divided by the $\rm 22~\mu m$ flux within FWHM of $^{12}$CO $J=1-0$ observation, both subtract the flux within nuclear region. Mark with ``$\star\star$'' uses scaling factor derived from $\rm H\alpha$ images (See Sect.~\ref{subsec:Intensityandmass} for details.). (3) $\rm M_{H_2,disk}^o$ is the original molecular gas mass from all the IRAM observed positions except nuclear region. (4) $\rm M_{H_2,disk}$ is the corrected molecular gas mass, calculated by multiplying $\rm M_{H_2,disk}^o$ by $\textit{f}_{\text{IR}}$. (5) $\rm M_{H_2,nucleus}^o$ is the original molecular mass from nuclear region. Those original molecular mass estimate from $^{12}$CO $J=1-0$ and assume the filling factor $f$ within the beam as $\propto~f^{-1}$. (6) $\rm M_{H_2}$ is the total molecular gas mass we use in the Sect.~\ref{sec:discussion} and derived by summing the $\rm M_{H_2,disk}$ and $\rm M_{H_2,nucleus}^o$, except for the value marked with ``$\star$''. Mark with ``$\star$'' is just obtained from original data of $\rm M_{H_2,nucleus}^o$ and $\rm M_{H_2,disk}^o$. (7) $\rm M_{H_2,^{13}CO}$ is the total molecular gas mass derived from $^{13}$CO $J=1-0$ emission with the same method and $\textit{f}_{\text{IR}}$ as for $^{12}$CO $J=1-0$. (8) $\rm \Sigma_{M_{H_2}}$ is the total molecular gas surface density calculated using the $\rm 22~\mu m$ diameter. The uncertainties are limited to the statistical errors derived from the measurements.

\end{table*}

\begin{figure}
\begin{center}
\includegraphics[width=0.47\textwidth]{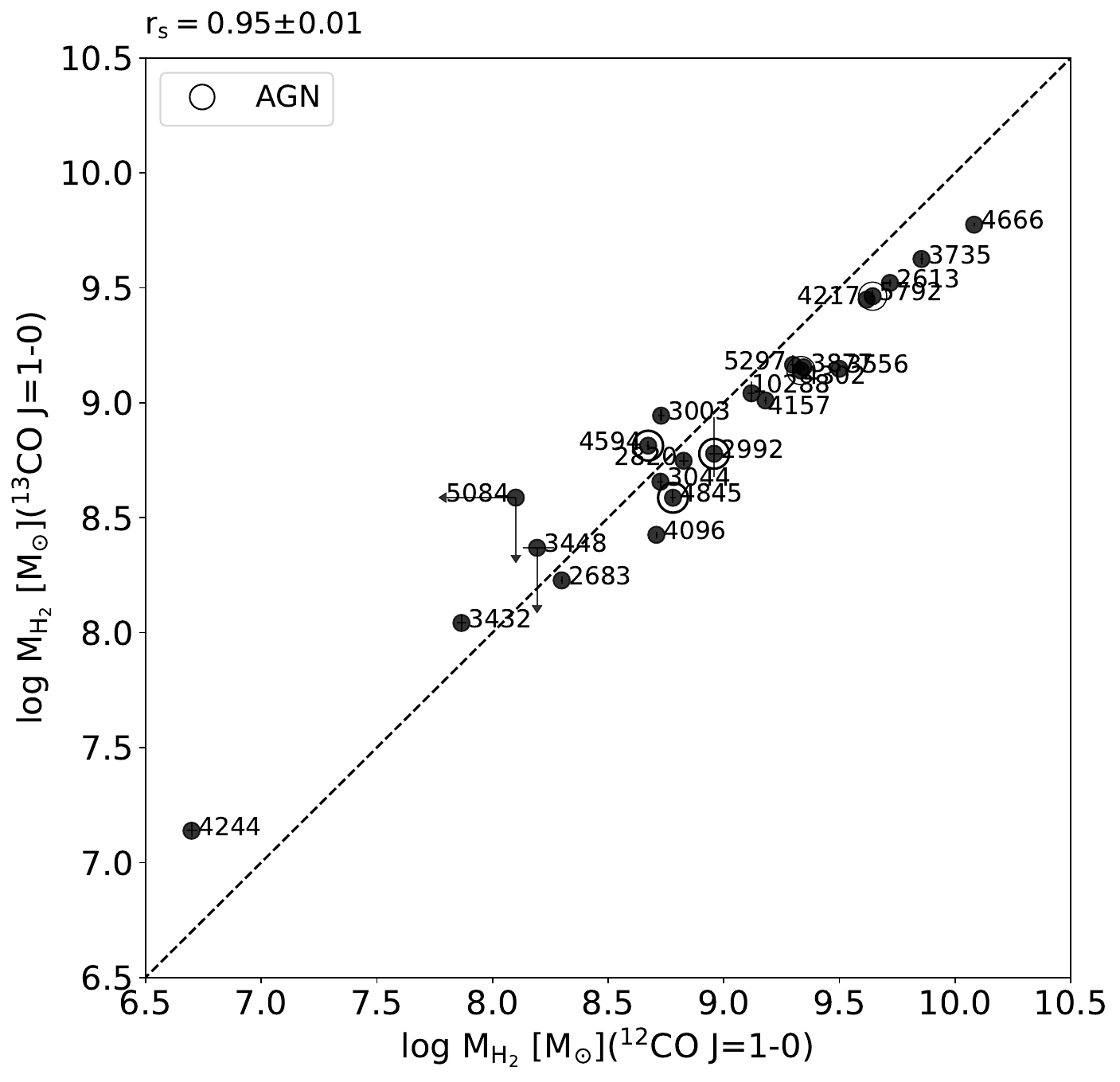}
\caption{
Comparison of the total molecular gas mass derived from the $^{12}$CO $J=1-0$ and $^{13}$CO $J=1-0$ lines for 23 galaxies in our sample. The dashed line represents the 1:1 ratio. Galaxies hosting ERB (extremely radio bright) AGNs, which may significantly influence the measurements of radio flux density, are marked with a thick black circle, while those with RB (radio bright) AGNs, less likely to strongly affect the radio flux measurements, are marked with a thin black circle. These AGNs are labeled as 'ERB' and 'RB,' respectively, in the last column of Table~1 in \citet{Li16a}. 
}\label{fig:12mass213mass}
\end{center}
\end{figure}

For comparison, we also derive the total molecular gas mass from the $^{13}$CO $J=1-0$ emission using the same method and scaling factor as for $^{12}$CO $J=1-0$. The results are listed in Table~\ref{table:masstable}. We compare the two masses in Fig.~\ref{fig:12mass213mass}, and find that for most of the galaxies these two masses show a tight linear correlation, but the molecular gas mass derived from $^{13}$CO $J=1-0$ may be systematically smaller, with a median value of $\rm M_{H_2,^{13}co}$/$\rm M_{H_2,^{12}co}$ ratio of $0.70\pm+0.08$ and a fitted slope of $0.77\pm0.03$ in log space. The discrepancies between the molecular gas mass derived from $^{12}$CO $J=1-0$ emission and those derived from $^{13}$CO $J=1-0$ emission are most likely due to the differences in the $X$ factor dependences on galaxies or environments as discussed above in this section, which is not considered in our calculations. For the high-mass end, there may be an overestimation of the $X_{\rm ^{12}CO(1-0)}$ factor, as evidenced by $\rm M_{H_2,^{12}co}$ being greater than $\rm M_{H_2,^{13}co}$ in the galaxy disk, and in the nucleus regions where $\rm M_{H_2,^{12}co,nucleus}$ also exceeds $\rm M_{H_2,^{13}co,nucleus}$. Moreover, the slightly higher critical density for excitation of $^{13}$CO $J=1-0$ emission ($\sim 600\rm~cm^{-3}$; \citealt{Jim17,Cormier18}) compared to $^{12}$CO $J=1-0$ makes it weaker or undetectable in more diffuse regime, particularly across most galactic disks in our sample. Compared with resolved observations from ALMA (Atacama Large Millimetre/submillimetre Array), such as NGC~2992 \citep{Zanchettin23} and NGC~4594 \citep{Sutter22}, the total molecular gas mass derived using the same $X$ factor shows no significant variation. In the subsequent discussions, we will only adopt the molecular gas mass derived from the $^{12}$CO $J=1-0$ line.

\subsection{CO line ratios and physical conditions of molecular gas}\label{subsec:RatioPara}

\begin{figure}
\begin{center}
\includegraphics[width=0.47\textwidth]{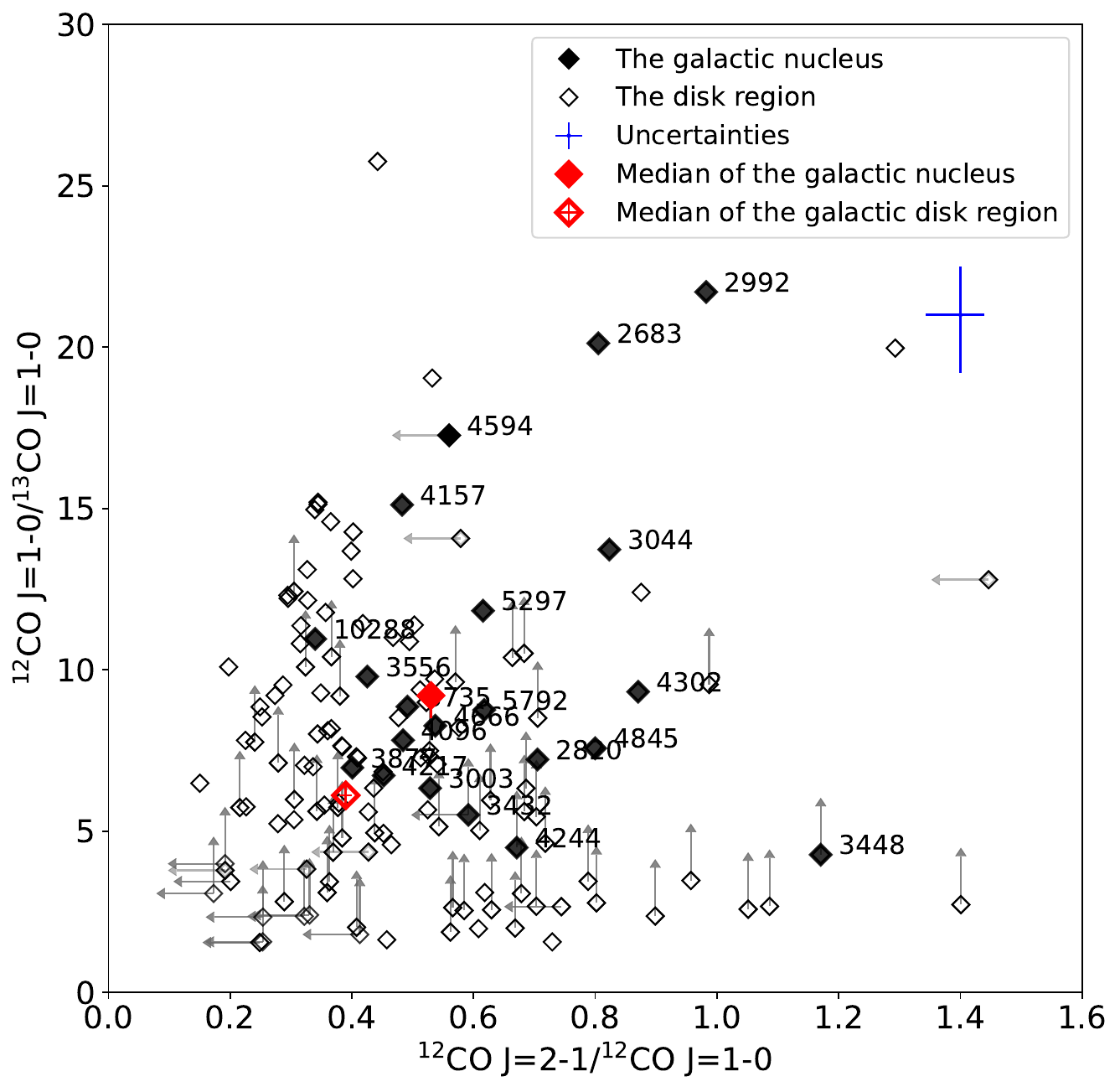}
\caption{A comparison of the $^{12}$CO/$^{13}$CO~$J=1-0$ and $^{12}$CO~$J=2-1$/$J=1-0$ line ratios for 23 galaxies in our sample. The disk regions and galactic nuclei are represented with open and filled symbols, respectively. The blue lines show the mean error for points without upper or lower limits. The red symbols represent the median value for the sample.
}\label{fig:ratio2ratio}
\end{center}
\end{figure}

In this section, we present the measured $^{12}$CO/$^{13}$CO~$J=1-0$ and $^{12}$CO~$J=2-1$/$J=1-0$ line ratios (Fig.~\ref{fig:ratio2ratio}), and derive some physical parameters based on them.

The $^{12}$CO/$^{13}$CO~$J=1-0$ ratio can be affected by both the optical depth and the isotopic abundance of the molecular cloud. In regions with active SF, the $^{12}$CO emission is enhanced in warm, diffuse clumps due to a significant decrease in the optical depth of the $^{12}$CO-rich envelopes, which allows more $^{12}$CO photons to escape. In colder molecular clouds, the abundance of $^{13}$CO increases primarily through isotope exchange reactions. Additionally, $^{13}$C can be produced during the CNO cycle in intermediate-mass stars' envelopes \citep{Truran77}. These combined factors in environments with intense SF can lead to an elevated $^{12}$CO/$^{13}$CO~$J=1-0$ line ratio (e.g., \citealt{Tan11, Jim17, Cormier18}).
In our sample, the $^{12}$CO/$^{13}$CO~$J=1-0$ ratio in the galactic nucleus is slightly higher than that in the galactic disk, and also slightly higher than the ratio observed in the Milky Way's center \citep{Solomon79}. This may be due to the influence of a few galaxies with particularly intense star formation activity in their nuclear regions.
The median value of the $^{12}$CO/$^{13}$CO~$J=1-0$ ratio for the galactic nucleus is $8.6^{+0.4}_{-0.9}$ with a typical range of 6-23. In contrast, the median value for the galactic disk is $6.1\pm0.4$, with a typical range of 1-19. Both single-pointing (e.g., \citealt{Aalto95, Crocker12}) and spatially resolved observations of nearby galaxies (e.g., \citealt{Paglione01, Cao17,Cao23}) show a clear variation of the $^{12}$CO/$^{13}$CO~$J=1-0$ ratio either between different galaxies or across individual galaxies. For normal late-type galaxies, the ratio typically varies from 5 to 15 (e.g., \citealt{Meier04,vila15,Roman16}), consistent with the measurements of our sample.

The $^{12}$CO~$J=2-1$/$J=1-0$ ratio is primarily determined by the temperature and optical depth. It reflects the structure and heating sources of the molecular clouds. Gas with a low $^{12}$CO~$J=2-1$/$J=1-0$ ratio $\textless$~0.7 typically originates from the extended low-density envelopes of molecular clumps, while the high ratio in the range of 0.7-1 is also primarily influenced by optical depth \citep{Hasegawa97} and tends to arise from highly concentrated molecular clumps characterized by steep density gradients and thin CO-emitting envelopes (e.g., \citealt{Penaloza17}). For very high $^{12}$CO~$J=2-1$/$J=1-0$ ratios ($\textgreater$ 1), the local environment has a significant influence, such as UV photons from young stars, shock waves from supernova explosions \citep{Hasegawa97}, or through effective heating mechanisms in extreme starburst environments (\citealt{Papadopoulos12}), including factors like supersonic turbulence of the molecular clouds, cosmic ray ionization rates, and the interstellar radiation field (e.g., \citealt{Penaloza18}).

As shown in Fig.~\ref{fig:ratio2ratio}, the median value of the $^{12}$CO~$J=2-1$/$J=1-0$ line ratio for the galactic nucleus is $0.53^{+0.04}_{-0.02}$, with a typical range of 0.5-1.4. The median value of this ratio in the galactic disk is $0.39\pm0.01$ with a range of 0.2-1.5. These ranges are in general comparable to measurements of other nearby galaxies (e.g., \citealt{Leroy22}). Although the galactic nucleus is often expected to have a higher $^{12}$CO~$J=2-1$/$J=1-0$ than the galactic disks (e.g., \citealt{Leroy09,Brok21,Yajima21}), this difference is not very significant for our sample, except for a few extreme cases. This indicates the low SF activity of most of our sample galaxies.

\begin{figure}
\begin{center}
\includegraphics[width=0.47\textwidth]{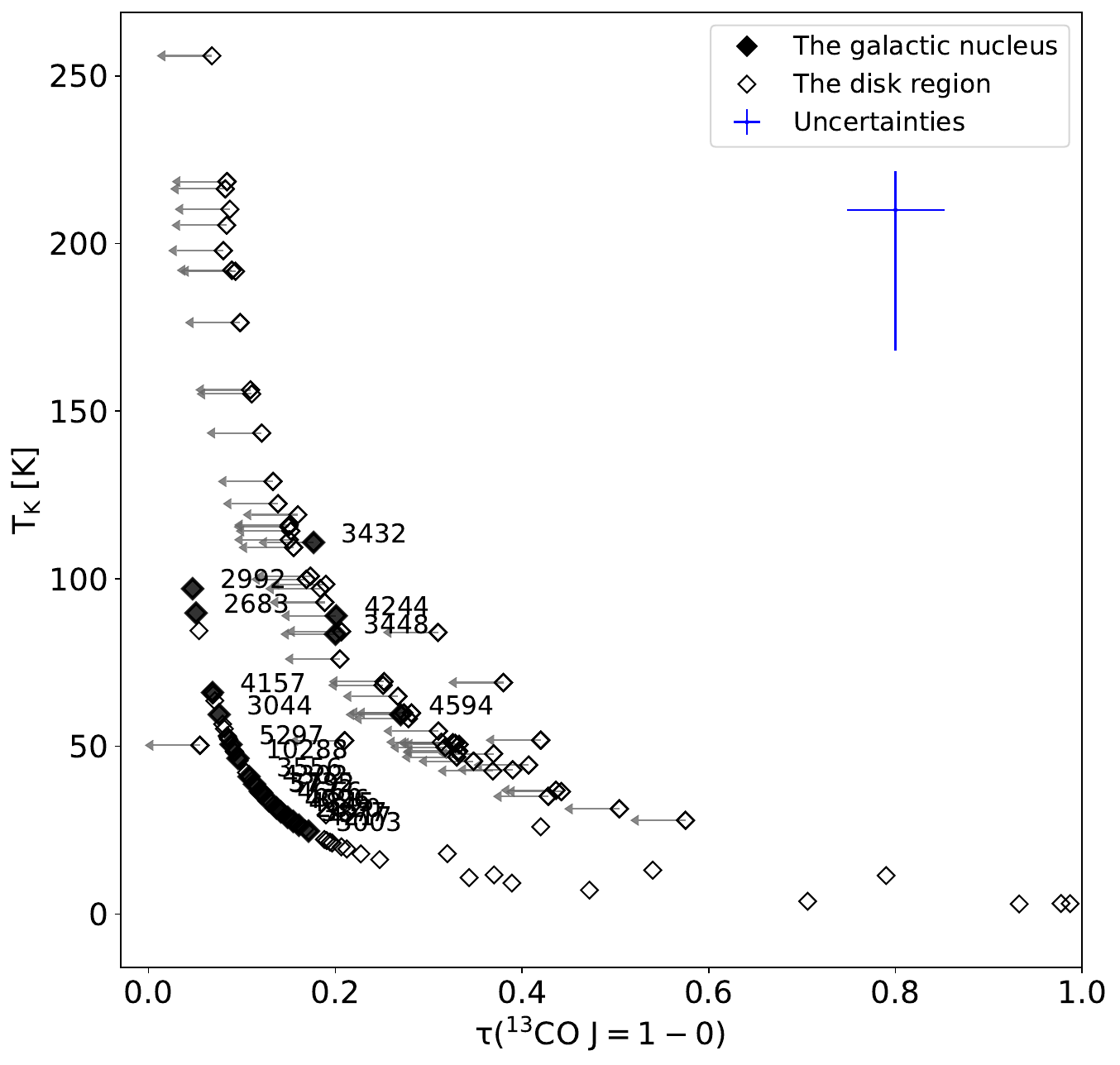}
\caption{The kinetic temperature ($\rm T_{\rm K}$) shows as a function of the optical depth of $^{13}$CO ($\tau_{\rm ^{13}CO}$), both derived from the nuclear and disk regions of the molecular gas under the assumption of LTE. The blue lines indicate mean error.
}\label{fig:tau2Tk}
\end{center}
\end{figure}

Under LTE conditions and some assumptions in Sect.~\ref{subsec:DataReduction}, the distributions of $\tau_{\rm ^{13}CO}$ and $\rm T_{K}$ are presented in Fig.~\ref{fig:tau2Tk}. The equations used are the same as in Paper I and are given by:
\begin{equation}\label{equtau}
\tau(\rm ^{13}CO)=-\ln[1-\frac{I_{\rm^{13}CO~(J=1-0)}}{I_{\rm ^{12}CO~(J=1-0)}}],
\end{equation}
where the $\rm I_{\rm^{12}CO}$ and $\rm~I_{\rm^{13}CO}$ are the integrated intensity of emission lines, and
\begin{equation}\label{equtk}
\rm\frac{N_{\rm H_2}}{\rm cm^{-2}}=2.25\times10^{20}[\frac{\tau_{\rm ^{13}CO}}{1-e^{-\tau_{\rm ^{13}CO}}}]\frac{I_{\rm ^{13}CO~J=1-0}}{1-e^{-5.29/T_{\rm ex}}},
\end{equation}
where the excitation temperature ($\rm~T_{\rm ex}$), under LTE, can be approximated as the kinetic temperature ($\rm~T_{\rm K}$). The $\tau_{\rm ^{13}CO}$ in the nuclear regions of our sample galaxies is typically $<0.25$ and has a median value of $0.09\pm0.01$. This value is generally higher than the $\tau_{\rm ^{13}CO}$ of nuclear starburst galaxies such as M82 (e.g., $\tau_{\rm ^{13}CO}\sim0.07$; \citealt{Tan11}). The median value of the $\rm T_{K}$ in the nuclear region is $40.9^{+8.9}_{-2.1}$~K, with a typical range of 3-256. In a few galaxies with nuclear activities (e.g., NGC~2992 and NGC~3448), this value can be as high as $\sim100\rm~K$. In the nuclear region of these galaxies, the $^{12}$CO $J=2-1$/$J=1-0$ ratio is often $\gtrsim1$, further suggesting the possibility additional heating from the AGN. 

\section{Discussion}\label{sec:discussion}

In this section, we will compare our measured molecular gas properties and the atomic gas properties collected from archive to other galaxy properties, in order to understand the role of cold gas in galaxy evolution. In the follow-up correlation analysis, we will use Spearman's rank order coefficient ($r_s$; by definition $-1~<~r_s~<1$) to describe the tightness of the correlation. We consider a tight correlation with $|r_s|~>~0.6$, a weak correlation with $0.6~>~|r_s|~>~0.3$, and no correlation with $|r_s|~<~0.3$ (e.g., \citealt{Li13a,Li13b}). Since there is no firm detection of CO $J=1-0$ in NGC~5084, this galaxy is excluded from subsequent analyses and only shows in Fig.~\ref{fig:mass2mass}.

\subsection{Molecular to atomic gas mass ratio}\label{subsec:mass2mass}

\subsubsection{Molecular and atomic gas content of the sample galaxies}\label{subsubsec:HIH2mass}

There are many \ion{H}{I} surveys of nearby galaxies reaching different detection limits and angular resolutions. For example, the THINGS survey conducted with the VLA reaches a typical \ion{H}{I} column density detection limit of $N_{\rm~HI}\sim10^{19}\rm~cm^{-2}$ and an angular resolution of $30^{\prime\prime}$ (e.g., \citealt{Walter08}). The HALOGAS survey conducted with the Westerbork Synthesis Radio Telescope (WSRT), typically with $\sim17$ times the exposure time of THINGS, reaches a detection limit of $N_{\rm~HI}\sim10^{19}\rm~cm^{-2}$ and an angular resolution of 20 $^{\prime\prime}$ - 30 $^{\prime\prime}$ (e.g., \citealt{Heald11b}). The FEASTS survey conducted with the single-dish Five-hundred-meter Aperture Spherical Telescope (FAST), without missing flux, reaches a typical detection limit of $N_{\rm~HI}\sim10^{17.7}\rm~cm^{-2}$ and an angular resolution of $3.2^\prime$ (e.g., \citealt{Wang23, Wang24}). We consider that the contribution from low column density \ion{H}{I} is not significant for our analysis, as the primary focus is on higher column densities that dominate the gas content in galaxies.

In this paper, since we only need the total \ion{H}{I} gas mass within the galactic disk, we directly collect the atomic gas mass ($\rm M_{HI}$) of our sample galaxies from different literatures, as summarized in Table~\ref{table:Parameters}. $\rm M_{HI}$ of 14 galaxies is obtained from the CHANG-ES VLA C-configuration \ion{H}{I} observations presented in \citet{Zheng22}. For the other galaxies, we typically only obtain the \ion{H}{I} flux measured with different telescopes (Table~\ref{table:Parameters}), and convert them to $\rm M_{HI}$ using the following relation:
\begin{equation}\label{MHI}
 \rm M_{HI}/M_\odot = 2.356 \times 10^5~S_{HI}~d^2,
\end{equation}
where $\rm S_{HI}$ is the integrated \ion{H}{I} line flux in unit of $\rm Jy~km~s^{-1}$, d is the distance to the galaxy in unit of Mpc (taken from \citealt{Vargas19}). The derived $\rm M_{HI}$ is also summarized in Table~\ref{table:Parameters}. 

\begin{figure}
\begin{center}
\includegraphics[width=0.47\textwidth]{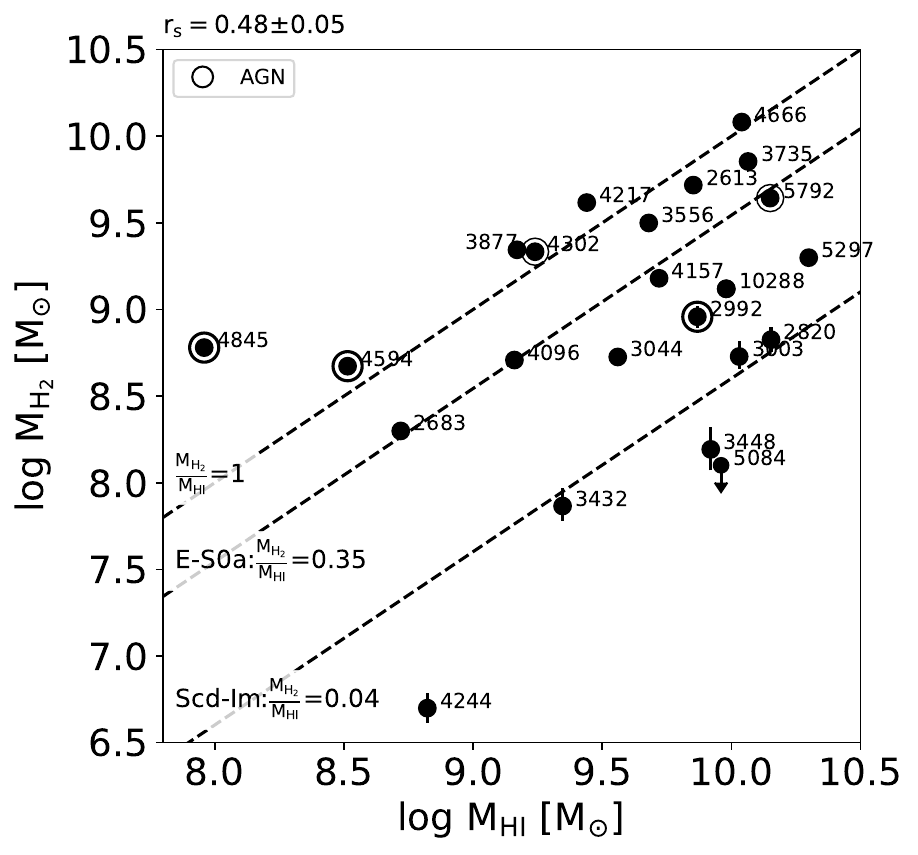}
\caption{$\rm M_{H_2}$ vs. $\rm M_{HI}$. The dotted line represents the $\rm M_{H_2}$/$\rm M_{HI}$ ratio, where the median value of 0.35 and 0.04 are corresponding to different galaxy types, as obtained from \citet{Lisenfeld11}. The circle symbols represent the AGN, as shown in Fig.~\ref{fig:12mass213mass}.
}\label{fig:mass2mass}
\end{center}
\end{figure}

The mass fraction of molecular and atomic gas and the transformation between them are crucial in the study of galaxy formation and evolution. Previous works (e.g., \citealt{Young89, Saintonge11, Catinella18, Lisenfeld19}) suggest a weak correlation between the mass of molecular and atomic gases in nearby galaxies. This weak correlation is also evident in our sample, as shown in Fig.~\ref{fig:mass2mass} ($r_s\approx0.49$). Despite the weak correlation, the scatter of $\rm M_{H_2}$ at a given $\rm M_{HI}$ is up to more than one order of magnitude, indicating a wide range of the molecular to atomic gas mass ratio, regardless of the non-uniformity of $\rm M_{HI}$ measurements from different literatures. The $\rm M_{H_2}$-$\rm M_{HI}$ relation, or the $\rm M_{H_2}$/$\rm M_{HI}$ ratio, could potentially be affected by the AGN activity and the morphological type of the galaxies, as suggested in many previous works (e.g., \citealt{Young89, Casoli98, Lisenfeld11}), or by some special cases. For example, there are two significant outliers in Fig.~\ref{fig:mass2mass}. NGC~4845, which has a very active AGN \citep{Irwin15}, does not show a deficiency in molecular gas mass compared to other galaxies, but rather has the lowest atomic gas content in our sample. NGC~4244 is a low surface brightness galaxy and is one of the latest type galaxies in the CHANG-ES sample; the low molecular gas content may be consistent with its low stellar mass and SFR (both lowest in the sample; \citealt{Li16a}). However, we do not find a significant trend showing such a dependence in Fig.~\ref{fig:mass2mass}. This is possibly because the CHANG-ES sample is selected without extreme AGN activities and all the galaxies are spiral, in a relatively narrow morphological type range.

\subsubsection{Gas mass ratio and stellar mass}\label{subsubsec:massratio2stellarmass}

In this subsection, we aim to further investigate the dependence of the molecular to atomic gas mass ratio ($\rm M_{H_2}$/$\rm M_{HI}$) on other galaxy parameters, such as the stellar mass ($\rm M_{\star}$).

\begin{figure}
\begin{center}
\includegraphics[width=0.47\textwidth]{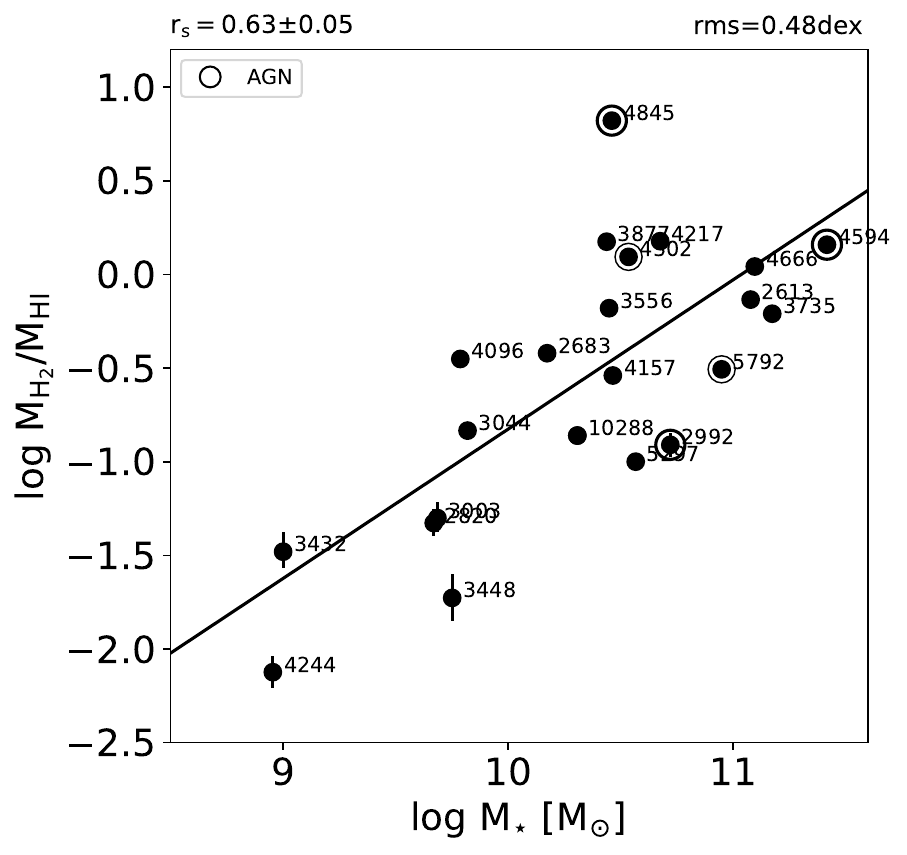}
\caption{The distribution of $\rm M_{H_2}$/$\rm M_{HI}$ ratio with $\rm M_{\star}$. The circle symbol of each galaxy are the same with Fig.~\ref{fig:12mass213mass}.
}\label{fig:massratio2stellar}
\end{center}
\end{figure}

Fig.~\ref{fig:massratio2stellar} shows that $\rm M_{H_2}$/$\rm M_{HI}$ is strongly correlated with $\rm M_{\star}$ ($r_s\approx0.63$) in our sample, consistent with previous studies from extensive samples (e.g., \citealt{leroy05, Saintonge11, Lisenfeld11, Bothwell14, Catinella18}). We have fitted this correlation with
\begin{equation}\label{massratio2mass}
\rm log~\frac{M_{H_2}}{M_{HI}}=(0.79\pm0.19)~log~(\frac{M_{\star}}{M_\odot})-(8.78\pm2.01),
\end{equation}
where $\rm M_{\star}$ is derived from the K-band luminosity and the mass-to-light ratio, as presented in \citet{Li16a}. This relation is primarily affected by the efficiency of atomic-to-molecular gas conversion, which depends on two main factors: metallicity and pressure. The increase of $\rm M_{\star}$ leads to higher metallicity and/or dust in the ISM, which shields the molecular gas from dissociation by the radiation field (e.g., \citealt{Elmegreen93, Fu10, Krumholz09}). Under hydrostatic equilibrium, stronger gravitational potential wells lead to higher pressures, compressing gas and increasing the column density of the gas, thereby enhancing the efficiency of atomic-to-molecular gas conversion (e.g., \citealt{Elmegreen89,Elmegreen93,Wong02,Leroy08}). The gravitational potential in galaxies can be dominated either by stars or by the self-gravity of gas, depending on the specific environment.

Galaxies with lower stellar masses, located in the lower left of Fig.~\ref{fig:massratio2stellar}, also appear in the lower right of Fig.~\ref{fig:mass2mass}, where their molecular gas masses are an order of magnitude lower than atomic gas mass. The primary reason is their relatively low stellar masses, which cause a shallow potential well, lower pressure, and lower gas surface density, hindering the conversion of atomic gas into molecular gas, especially in the lowest-mass galaxy NGC~4244. Additionally, galaxies with shallow potential wells may struggle to retain their metallicity (e.g., \citealt{Tremonti04}). The increase in the $\rm M_{H_2}$/$\rm M_{HI}$ ratio for NGC~4845 is mainly due to a deficiency in $\rm M_{HI}$.

\subsection{Star formation law}\label{subsec:SFR}

\begin{figure}
\begin{center}
\includegraphics[width=0.47\textwidth]{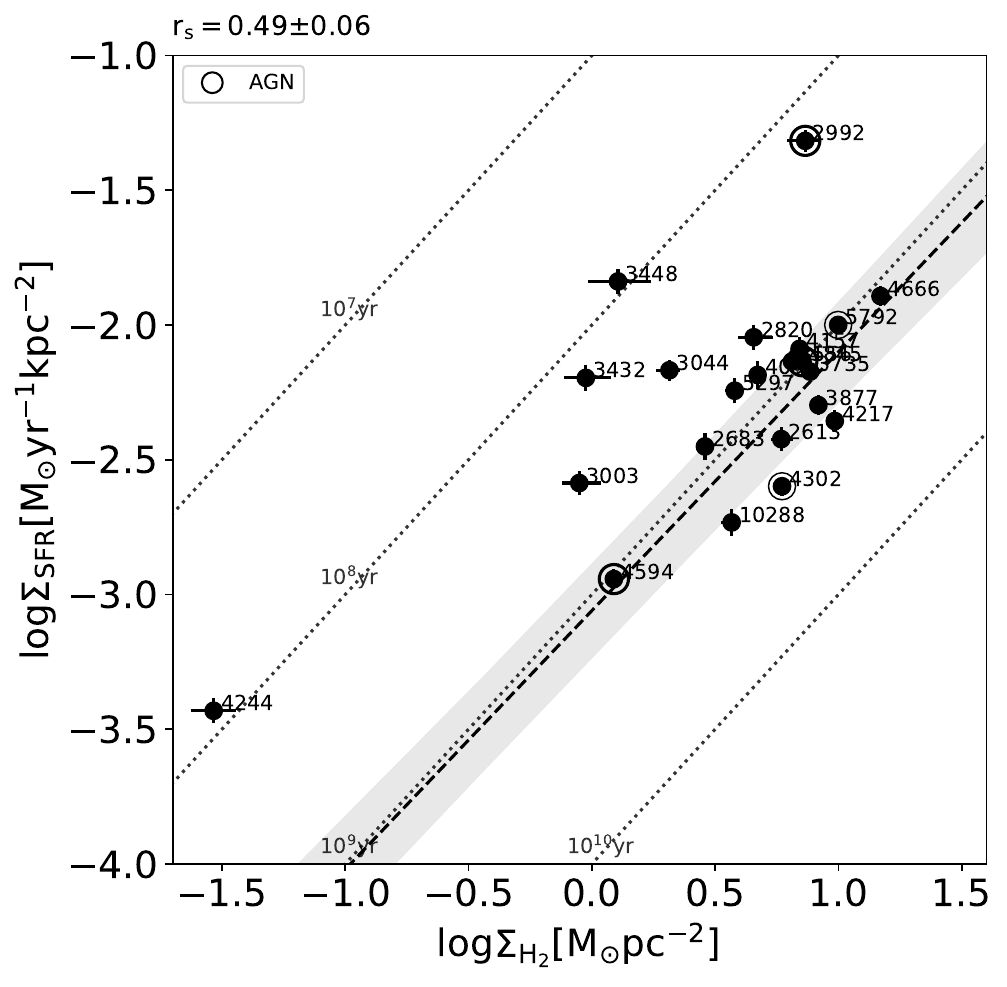}
\caption{The $\rm \Sigma_{H_2}$ is shown as a function of $\rm \Sigma_{SFR}$. The AGN symbol is the same as Fig.~\ref{fig:12mass213mass}. The dashed line represents the average fitting of seven galaxies in \citet{Bigiel08}, and the shaded area indicates the $1\sigma$ uncertainty of the fit. The dotted lines correspond to gas depletion times of $10^{10}$, $10^9$, $10^8$ and $10^7$ years from bottom to the top. 
}\label{fig:SFR}
\end{center}
\end{figure}

The cold atomic and molecular gas provide fuels to continue SF, with the latter directly connected to the current SF processes. The Kennicutt–Schmidt SF law describe the relationship between the surface density of the cold gas and the surface density of the SFR measured with different tracers (e.g., \citealt{Kennicutt98,Kennicutt12}). With spatially resolved measurement of the molecular gas distribution on the galactic disk, our sample is optimized to examine the SF law. We compare our sample to the well calibrated SF law based on similarly spatially resolved cold gas and SFR observations of nearby galaxies from \citet{Bigiel08} in Fig.~\ref{fig:SFR}. We also adopt the gas depletion timescale to describe how active is the SF in the galaxy. It is defined as the timescale to consume the SF fuel at the current SFR, or the inverse of the SF efficiency: $\rm\tau_{H_2}=M_{H_2}/SFR$. 

In Fig.~\ref{fig:SFR}, the SF law is presented in the form of $\Sigma_{\rm SFR}-\Sigma_{\rm H_2}$, where we calculate $\rm~\Sigma_{H_2}$ based on the total measured molecular gas mass $\rm M_{H_2}$ and the SF diameter of the galaxy measured with the $22\rm~\mu m$ data (obtained from \citealt{Wiegert15} and listed in Table~\ref{table:Parameters}). We directly adopt the disk-average SFR surface density $\Sigma_{\rm~SFR}$ computed base on the spatially resolved $22\rm~\mu m$ and H$\alpha$ data \citep{Vargas19}. In Fig.~\ref{fig:SFR}, we also plot the $\Sigma_{\rm SFR}-\Sigma_{\rm H_2}$ relationships corresponding to different gas depletion timescales. Our sample mostly falls on the best-fit relation from \citet{Bigiel08}, which corresponds to a depletion timescale of $\rm\tau_{H_2}\sim10^9\rm~yr$. This is consistent with some moderately SF galaxies (e.g., \citealt{Bigiel11, Leroy13, Lisenfeld19}), but could be one order of magnitude larger than extreme starburst galaxies, such as luminous infrared galaxies and ultraluminous infrared galaxies, which are know to have higher SF efficiencies and shorter gas depletion timescale (e.g., \citealt{Saintonge11}).

As show in Fig.~\ref{fig:SFR}, our sample shows a relatively large scatter compared to the usually tight correlation between $\Sigma_{\rm SFR}$ and $\Sigma_{\rm H_2}$ (e.g., \citealt{Kennicutt12}). This is mainly because of the relatively small dynamical range of both parameters in our sample, which mainly represent inactive or normal SF galaxies. Furthermore, there are a few significant outliers from the SF law, which enlarge this scatter. For example, NGC~4244 is a dwarf galaxy with small gas content and weak SF, and the IR and H$\alpha$ emissions used to calculate the SFR may be contaminated by the contributions from some old stellar components (e.g., \citealt{Vargas18,Vargas19}). The real uncertainty of the measured $\Sigma_{\rm SFR}$ and $\Sigma_{\rm H_2}$ should be larger than those presented in Fig.~\ref{fig:SFR}, which include only statistical errors. NGC~2992 has an extremely active AGN and relatively small angular size. The SFR determined largely from the WISE $22\rm~\mu m$ data should be largely overestimated, with contributions from the strong AGN emissions. \citet{Xu24} analyzed the stellar population in NGC~2992 and revealed that AGN outflows have suppressed SF in the nuclear region (<1~kpc), but trigger increased SF activity outside the nucleus. NGC~3448 also has an IR bright core, which may represent a weak AGN and cause the overestimate of the SFR, although high-resolution radio continuum observations suggest the peak of the radio emission may be off nucleus \citep{Irwin19}. NGC~3432 has a warped disk possibly impacted by the gravitational tidal force of the dwarf companions UGC~5983 and [KMK2013] LV J1052+3639. The galaxy may be in a stage when the tidal interaction helps triggering the SF on the galactic disk (e.g., \citealt{Renaud14,Pan18}). NGC~3003 exhibits unusually large \ion{H}{I}, H$\alpha$, and radio scales high compared to other CHANG-ES galaxies, which may be influence by the tidal interaction (e.g., \citealt{Krause18, Lu23}).

\subsection{Gas baryon budget}\label{subsec:barynoicTF}

\begin{figure}
\begin{center}
\includegraphics[width=0.47\textwidth]{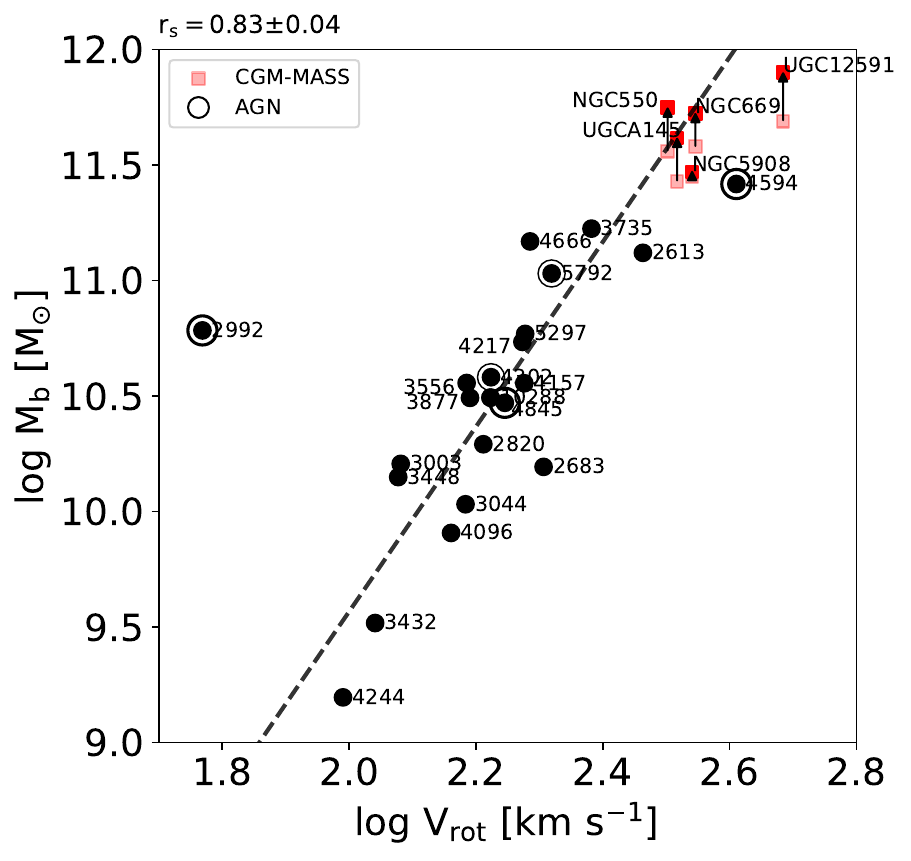}
\caption{The rotational velocity is shown as a function of the baryonic mass of galaxies. The AGN symbol is the same as Fig.~\ref{fig:12mass213mass}. The dashed line is the best-fit of baryonic Tully-Fisher relation from \citet{McGaugh05}. The red box symbols represent galaxies from the CGM-MASS sample, with lighter red boxes indicating those that only include stellar mass and darker red boxes adding the hot gas mass of the galaxy, both of which are obtained from \citet{Li17}. The mass variations for each galaxy, considering whether hot gas mass is included or not, are indicated by black arrows.
}\label{fig:TF}
\end{center}
\end{figure}

The fraction of baryons of a galaxy stored in different forms varies at different galaxy masses, and the gas in different phases could take a significant fraction of a galaxy's total baryon budget (e.g., \citealt{Werk14,Li17,Li18,Bregman18,Bregman22}). The baryonic Tully-Fisher relation is an empirical scaling relation between the baryonic mass $\rm M_b$ and the gravitation of the galaxy characterized with the rotation velocity $\rm V_{rot}$ (e.g., \citealt{McGaugh00}). In Fig.~\ref{fig:TF}, we plot our sample galaxies against the well-defined baryonic Tully-Fisher relation from \citet{McGaugh05}, using the sum of stellar mass and total gas (H$_2$ + \ion{H}{I}) masses, along with the rotation velocity obtained from archive (e.g., \citealt{Li16a}). In follow-up papers, we will further discuss the rotation curve of our sample galaxies measured based on both our CO data \citep{Li19,Jiang24} and some multi-wavelength data tracing different gas phases (e.g., \citealt{Zheng22,Li24}).

As expected, our sample galaxies distribute tightly around the best-fit baryonic Tully-Fisher relation from \citet{McGaugh05}. The most significant outlier is NGC~2992, where there is a strong AGN with broadened double peaked emission lines from the nuclear region (see an example from our CO observations in Fig.~\ref{fig:2992nucleus}) and a highly warped disk. The measured $\rm V_{rot}$ could be largely biased and not related to the gravitation of the galaxy. The second largest outlier is NGC~4594, which is an early-type disk galaxy \citep{Jiang24} with significantly lower stellar+cold gas mass than expected from the baryonic Tully-Fisher relation. NGC~4594 is also the most massive galaxy in our sample. In order to double check if there is a systematic bias of massive disk galaxies from the baryonic Tully-Fisher relation in the high mass end, we further compare our sample to the Circum-Galactic Medium of MASsive Spirals (CGM-MASS) sample including the most massive spiral galaxies in the local Universe (Fig.~\ref{fig:TF}; \citealt{Li16b,Li17,Li18}). In these massive galaxies, the cold gas is poor, constituting approximately 10\% of the stellar mass \citep{Lisenfeld23}. In Fig.~\ref{fig:TF}, the light red symbols represent galaxies where only the stellar mass is considered; they indeed show a systematic bias to lower baryonic mass from the baryonic Tully-Fisher relation. Even when cold gas mass is included, this bias cannot be fully corrected, indicating either a larger halo mass (e.g., expected from the turn over of the stellar mass-halo mass relation; \citealt{Behroozi10}) or a lower fraction of baryonic mass converted to stars. However, these massive galaxies have a higher virial temperature ($\rm~T>10^6~K$), so the gravitation potential is deep enough to heat the gas to a temperature above the peak of the radiative cooling curve, converting it to X-ray emitting hot gas (e.g., \citealt{Li18}). The CGM-MASS sample has high quality XMM-Newton observations which could be used to measure the properties of the hot CGM. With the mass of this extended hot gas component included, these massive spirals fall closer to the baryonic Tully-Fisher relation (Fig.~\ref{fig:TF}). 

\begin{figure}
\begin{center}
\includegraphics[width=0.47\textwidth]{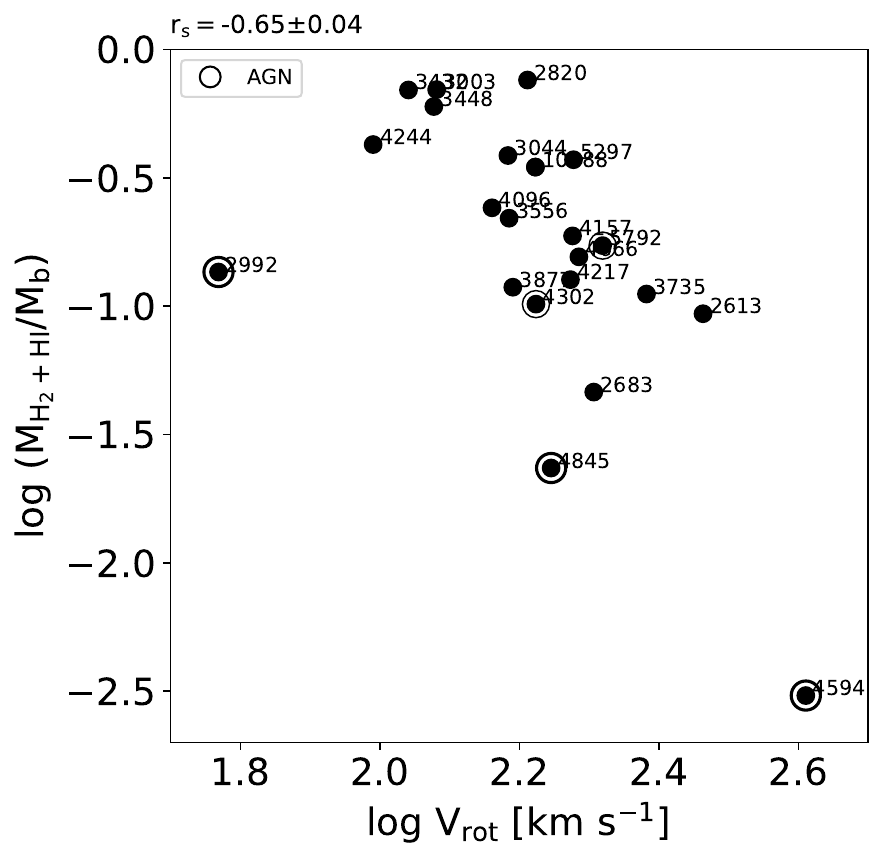}
\caption{The rotational velocity show as a function of the total cold gas mass to baryonic mass ratio of galaxies. The AGN symbol are the same with Fig.~\ref{fig:12mass213mass}. 
}\label{fig:GasMassFracVrot}
\end{center}
\end{figure}

As discussed above, the multi-phase gas could account for take a significant fraction of the baryon budget of a galaxy. We further examine the dependence of the cold (molecular plus atomic) gas to the total baryon mass ratio ($\rm M_{\rm H_2+HI}/M_{\rm b}$) on the gravitational potential of a galaxy traced by $\rm V_{rot}$ in Fig.~\ref{fig:GasMassFracVrot}. Excluding NGC~2992 whose $\rm V_{rot}$ may not be a good tracer of the gravitational potential, the other galaxies show a fairly good anti-correction between $\rm M_{H_2+HI}/M_{\rm b}$ and $\rm V_{rot}$. This indicates that more massive spiral galaxies tend to be poorer in cold gas, resulting in their quiescence in SF (e.g., \citealt{Salim07}). However, these galaxies do not necessarily to be gas poor, as a significant fraction of the gas could be heated to an X-ray emitting temperature, resulting in an overall positive correlation between the hot gas mass fraction and the gravitational potential (e.g., \citealt{Li14,Li17,Li18}). The radiative cooling of the hot CGM may also partially compensate the gas consumed in SF, especially for relatively quiescent and massive galaxies (e.g., \citealt{Li14,Li17}). In conclusion, although most of our sample galaxies fall on the well defined baryonic Tully-Fisher relation, some most massive spirals apparently appear depleted in cold gas. The depletion could at least be partially explained that a significant fraction of the gas in the halo has been heated to an X-ray emitting temperature above the peak of the radiative cooling curve.

\section{Summary and Conclusion}\label{sec:summary}

We present IRAM 30m observations of the $^{12}$CO~$J=1-0$, $^{13}$CO~$J=1-0$, and $^{12}$CO~$J=2-1$ lines along the disk of 23 nearby spiral galaxies selected from the CHANG-ES sample. We then calculate the ratios between different CO lines and derive the temperature and optical depth of the molecular gas. We further examine the dependence of the gas properties on other galaxy properties. Our main results and conclusions are summarized below:

$\bullet$ The $^{12}$CO~$J=1-0$, $^{13}$CO~$J=1-0$, and $^{12}$CO~$J=2-1$ lines are detected at 77\%, 40\%, and 73\% of the observed positions along the galactic disks, and 96\%, 78\%, and 96\% of the nuclear regions of the galaxies.

$\bullet$ Using the WISE $22\rm~\mu m$ image, we rescale the molecular gas mass measured from our CO line observations of individual regions to the total molecular gas mass of each galaxy. For most galaxies, the molecular gas mass derived from $^{13}$CO~$J=1-0$ is tightly correlated with mass derived from $^{13}$CO~$J=1-0$, but systematically lower than that from $^{12}$CO~$J=1-0$, possibly caused by the uncertainties of $X$ factor and lower detection rate of the weaker $^{13}$CO~$J=1-0$ line in the galaxy disk.

$\bullet$ After correcting for the beam dilution, the median values of the $^{12}$CO/$^{13}$CO~$J=1-0$ and $^{12}$CO~$J=2-1$/$J=1-0$ ratios in the nucleus/disk of galaxies are $\approx$8.6/6.1 and 0.53/0.39, respectively. The median values of the derived kinetic temperature and optical depth under the LTE assumption in the galaxy are $\approx 41\rm~K$ and 0.1, respectively. These values are all typical for non-starburst spiral galaxies.

$\bullet$ There is a weak correlation between the total molecular and atomic gas masses of our sample galaxies, but the molecular to atomic gas mass ratio $\rm M_{\rm H_2}/M_{\rm HI}$ shows a large scatter of more than one order of magnitude. $\rm M_{\rm H_2}/M_{\rm HI}$ could be affected by many factors. One of the most important ones appears to be the stellar mass of the galaxy, which shows a tight, nearly linear relation with $\rm M_{\rm H_2}/M_{\rm HI}$. This correlation indicates that in galaxies with higher stellar masses, increased metallicity and stronger gravitational potential accelerate the conversion of atomic gas into molecular gas, leading to a higher $\rm M_{\rm H_2}/M_{\rm HI}$ ratio. In our sample, galaxies with lower stellar masses also exhibit reduced $\rm M_{\rm H_2}/M_{\rm HI}$ ratios, with the molecular gas masses in these galaxies being an order of magnitude lower than the atomic gas masses.

$\bullet$ Most of our sample galaxies fall close to the Kennicutt–Schmidt SF law, with a median gas depletion timescale of $\sim1\rm~Gyr$. Some galaxies show significantly shorter gas depletion timescales of $\sim0.1\rm~Gyr$, and could be affected by various measurement uncertainties, as well as some special processes prompting the SF efficiency.

$\bullet$ Our sample galaxies fall on the baryonic Tully-Fisher relation, except for a few most massive galaxies, which have systematically lower baryonic mass at a given rotation velocity. The missing baryonic mass of these massive spirals is largely recovered by the inclusion of the extended hot CGM. The cold gas fraction of the total baryon mass decreases with increasing gravitational potential, because the gas is consumed either in early SF or being heated to the hot phase.

\bigskip
\noindent\textbf{\uppercase{acknowledgements}}
\smallskip\\
\noindent The authors would like to commemorate the late Prof. Yu Gao and express their deep gratitude for his invaluable guidance in observations and learning. The authors also acknowledge Dr. Zhi-Yu Zhang and Dr. Jian Fu for helpful discussions. J.T.L. acknowledges the financial support from the National Science Foundation of China (NSFC) through the grants 12273111, and also the science research grants from the China Manned Space Project. Q.H.T. acknowledges the support by the NSFC grant No. 12033004.


\bibliography{reference}{}

\begin{thebibliography}{116}
\expandafter\ifx\csname natexlab\endcsname\relax\def\natexlab#1{#1}\fi

\bibitem[{{Aalto} {et~al.}(1995){Aalto}, {Booth}, {Black}, \& {Johansson}}]{Aalto95}
{Aalto}, S., {Booth}, R.~S., {Black}, J.~H., \& {Johansson}, L.~E.~B. 1995, \aap, 300, 369

\bibitem[{{Aniano} {et~al.}(2011){Aniano}, {Draine}, {Gordon}, \& {Sandstrom}}]{Aniano11}
{Aniano}, G., {Draine}, B.~T., {Gordon}, K.~D., \& {Sandstrom}, K. 2011, \pasp, 123, 1218

\bibitem[{{Behroozi} {et~al.}(2010){Behroozi}, {Conroy}, \& {Wechsler}}]{Behroozi10}
{Behroozi}, P.~S., {Conroy}, C., \& {Wechsler}, R.~H. 2010, \apj, 717, 379

\bibitem[{{Bigiel} {et~al.}(2008){Bigiel}, {Leroy}, {Walter}, {Brinks}, {de Blok}, {Madore}, \& {Thornley}}]{Bigiel08}
{Bigiel}, F., {Leroy}, A., {Walter}, F., {et~al.} 2008, \aj, 136, 2846

\bibitem[{{Bigiel} {et~al.}(2011){Bigiel}, {Leroy}, {Walter}, {Brinks}, {de Blok}, {Kramer}, {Rix}, {Schruba}, {Schuster}, {Usero}, \& {Wiesemeyer}}]{Bigiel11}
{Bigiel}, F., {Leroy}, A.~K., {Walter}, F., {et~al.} 2011, \apjl, 730, L13

\bibitem[{{Bolatto} {et~al.}(2013){Bolatto}, {Wolfire}, \& {Leroy}}]{Bolatto13}
{Bolatto}, A.~D., {Wolfire}, M., \& {Leroy}, A.~K. 2013, \araa, 51, 207

\bibitem[{{Boselli} {et~al.}(2014){Boselli}, {Cortese}, {Boquien}, {Boissier}, {Catinella}, {Lagos}, \& {Saintonge}}]{Boselli14}
{Boselli}, A., {Cortese}, L., {Boquien}, M., {et~al.} 2014, \aap, 564, A66

\bibitem[{{Bothwell} {et~al.}(2014){Bothwell}, {Wagg}, {Cicone}, {Maiolino}, {M{\o}ller}, {Aravena}, {De Breuck}, {Peng}, {Espada}, {Hodge}, {Impellizzeri}, {Mart{\'\i}n}, {Riechers}, \& {Walter}}]{Bothwell14}
{Bothwell}, M.~S., {Wagg}, J., {Cicone}, C., {et~al.} 2014, \mnras, 445, 2599

\bibitem[{{Bregman} {et~al.}(2018){Bregman}, {Anderson}, {Miller}, {Hodges-Kluck}, {Dai}, {Li}, {Li}, \& {Qu}}]{Bregman18}
{Bregman}, J.~N., {Anderson}, M.~E., {Miller}, M.~J., {et~al.} 2018, \apj, 862, 3

\bibitem[{{Bregman} {et~al.}(2022){Bregman}, {Hodges-Kluck}, {Qu}, {Pratt}, {Li}, \& {Yun}}]{Bregman22}
{Bregman}, J.~N., {Hodges-Kluck}, E., {Qu}, Z., {et~al.} 2022, \apj, 928, 14

\bibitem[{{Cao} {et~al.}(2023){Cao}, {Wong}, {Bolatto}, {Leroy}, {Rosolowsky}, {Utomo}, {S{\'a}nchez}, {Barrera-Ballesteros}, {Levy}, {Colombo}, {Blitz}, {Vogel}, {Puschnig}, {Villanueva}, \& {Rubio}}]{Cao23}
{Cao}, Y., {Wong}, T., {Bolatto}, A.~D., {et~al.} 2023, \apjs, 268, 3

\bibitem[{{Cao} {et~al.}(2017){Cao}, {Wong}, {Xue}, {Bolatto}, {Blitz}, {Vogel}, {Leroy}, \& {Rosolowsky}}]{Cao17}
{Cao}, Y., {Wong}, T., {Xue}, R., {et~al.} 2017, \apj, 847, 33

\bibitem[{{Carter} {et~al.}(2012){Carter}, {Lazareff}, {Maier}, {Chenu}, {Fontana}, {Bortolotti}, {Boucher}, {Navarrini}, {Blanchet}, {Greve}, {John}, {Kramer}, {Morel}, {Navarro}, {Pe{\~n}alver}, {Schuster}, \& {Thum}}]{Carter12}
{Carter}, M., {Lazareff}, B., {Maier}, D., {et~al.} 2012, \aap, 538, A89

\bibitem[{{Casoli} {et~al.}(1998){Casoli}, {Sauty}, {Gerin}, {Boselli}, {Fouque}, {Braine}, {Gavazzi}, {Lequeux}, \& {Dickey}}]{Casoli98}
{Casoli}, F., {Sauty}, S., {Gerin}, M., {et~al.} 1998, \aap, 331, 451

\bibitem[{{Catinella} {et~al.}(2018){Catinella}, {Saintonge}, {Janowiecki}, {Cortese}, {Dav{\'e}}, {Lemonias}, {Cooper}, {Schiminovich}, {Hummels}, {Fabello}, {Ger{\'e}b}, {Kilborn}, \& {Wang}}]{Catinella18}
{Catinella}, B., {Saintonge}, A., {Janowiecki}, S., {et~al.} 2018, \mnras, 476, 875

\bibitem[{{Chaves} \& {Irwin}(2001)}]{Chaves01}
{Chaves}, T.~A. \& {Irwin}, J.~A. 2001, \apj, 557, 646

\bibitem[{{Cicone} {et~al.}(2017){Cicone}, {Bothwell}, {Wagg}, {M{\o}ller}, {De Breuck}, {Zhang}, {Mart{\'\i}n}, {Maiolino}, {Severgnini}, {Aravena}, {Belfiore}, {Espada}, {Fl{\"u}tsch}, {Impellizzeri}, {Peng}, {Raj}, {Ram{\'\i}rez-Olivencia}, {Riechers}, \& {Schawinski}}]{Cicone17}
{Cicone}, C., {Bothwell}, M., {Wagg}, J., {et~al.} 2017, \aap, 604, A53

\bibitem[{{Cormier} {et~al.}(2018){Cormier}, {Bigiel}, {Jim{\'e}nez-Donaire}, {Leroy}, {Gallagher}, {Usero}, {Sandstrom}, {Bolatto}, {Hughes}, {Kramer}, {Krumholz}, {Meier}, {Murphy}, {Pety}, {Rosolowsky}, {Schinnerer}, {Schruba}, {Sliwa}, \& {Walter}}]{Cormier18}
{Cormier}, D., {Bigiel}, F., {Jim{\'e}nez-Donaire}, M.~J., {et~al.} 2018, \mnras, 475, 3909

\bibitem[{{Courtois} \& {Tully}(2015)}]{Courtois15}
{Courtois}, H.~M. \& {Tully}, R.~B. 2015, \mnras, 447, 1531

\bibitem[{{Crocker} {et~al.}(2012){Crocker}, {Krips}, {Bureau}, {Young}, {Davis}, {Bayet}, {Alatalo}, {Blitz}, {Bois}, {Bournaud}, {Cappellari}, {Davies}, {de Zeeuw}, {Duc}, {Emsellem}, {Khochfar}, {Krajnovi{\'c}}, {Kuntschner}, {Lablanche}, {McDermid}, {Morganti}, {Naab}, {Oosterloo}, {Sarzi}, {Scott}, {Serra}, \& {Weijmans}}]{Crocker12}
{Crocker}, A., {Krips}, M., {Bureau}, M., {et~al.} 2012, \mnras, 421, 1298

\bibitem[{{Dame} {et~al.}(2001){Dame}, {Hartmann}, \& {Thaddeus}}]{Dame01}
{Dame}, T.~M., {Hartmann}, D., \& {Thaddeus}, P. 2001, \apj, 547, 792

\bibitem[{{Davis} \& {Seaquist}(1983)}]{Davis83}
{Davis}, L.~E. \& {Seaquist}, E.~R. 1983, \apjs, 53, 269

\bibitem[{{de Vaucouleurs} {et~al.}(1991){de Vaucouleurs}, {de Vaucouleurs}, {Corwin}, {Buta}, {Paturel}, \& {Fouque}}]{deVaucouleurs91}
{de Vaucouleurs}, G., {de Vaucouleurs}, A., {Corwin}, Herold~G., J., {et~al.} 1991, {Third Reference Catalogue of Bright Galaxies}

\bibitem[{{den Brok} {et~al.}(2021){den Brok}, {Chatzigiannakis}, {Bigiel}, {Puschnig}, {Barnes}, {Leroy}, {Jim{\'e}nez-Donaire}, {Usero}, {Schinnerer}, {Rosolowsky}, {Faesi}, {Grasha}, {Hughes}, {Kruijssen}, {Liu}, {Neumann}, {Pety}, {Querejeta}, {Saito}, {Schruba}, \& {Stuber}}]{Brok21}
{den Brok}, J.~S., {Chatzigiannakis}, D., {Bigiel}, F., {et~al.} 2021, \mnras, 504, 3221

\bibitem[{{Elmegreen}(1989)}]{Elmegreen89}
{Elmegreen}, B.~G. 1989, \apj, 338, 178

\bibitem[{{Elmegreen}(1993)}]{Elmegreen93}
{Elmegreen}, B.~G. 1993, \apj, 411, 170

\bibitem[{{Evans}(1999)}]{Evans99}
{Evans}, Neal~J., I. 1999, \araa, 37, 311

\bibitem[{{Frerking} {et~al.}(1982){Frerking}, {Langer}, \& {Wilson}}]{Frerking82}
{Frerking}, M.~A., {Langer}, W.~D., \& {Wilson}, R.~W. 1982, \apj, 262, 590

\bibitem[{{Fu} {et~al.}(2010){Fu}, {Guo}, {Kauffmann}, \& {Krumholz}}]{Fu10}
{Fu}, J., {Guo}, Q., {Kauffmann}, G., \& {Krumholz}, M.~R. 2010, \mnras, 409, 515

\bibitem[{{Gilli} {et~al.}(2000){Gilli}, {Maiolino}, {Marconi}, {Risaliti}, {Dadina}, {Weaver}, \& {Colbert}}]{Gilli00}
{Gilli}, R., {Maiolino}, R., {Marconi}, A., {et~al.} 2000, \aap, 355, 485

\bibitem[{{Goldsmith} {et~al.}(2008){Goldsmith}, {Heyer}, {Narayanan}, {Snell}, {Li}, \& {Brunt}}]{Goldsmith08}
{Goldsmith}, P.~F., {Heyer}, M., {Narayanan}, G., {et~al.} 2008, \apj, 680, 428

\bibitem[{{Hasegawa}(1997)}]{Hasegawa97}
{Hasegawa}, T. 1997, in IAU Symposium, Vol. 170, IAU Symposium, ed. W.~B. {Latter}, S.~J.~E. {Radford}, P.~R. {Jewell}, J.~G. {Mangum}, \& J.~{Bally}, 39--46

\bibitem[{{Heald} {et~al.}(2011){Heald}, {J{\'o}zsa}, {Serra}, {Zschaechner}, {Rand}, {Fraternali}, {Oosterloo}, {Walterbos}, {J{\"u}tte}, \& {Gentile}}]{Heald11b}
{Heald}, G., {J{\'o}zsa}, G., {Serra}, P., {et~al.} 2011, \aap, 526, A118

\bibitem[{{Henry} \& {Worthey}(1999)}]{Henry99}
{Henry}, R.~B.~C. \& {Worthey}, G. 1999, \pasp, 111, 919

\bibitem[{{Hollenbach} \& {Thronson}(1987)}]{Hollenbach87}
{Hollenbach}, D.~J. \& {Thronson}, Harley~A., J., eds. 1987, Astrophysics and Space Science Library, Vol. 134, {Interstellar processes}

\bibitem[{{Huchra} {et~al.}(2012){Huchra}, {Macri}, {Masters}, {Jarrett}, {Berlind}, {Calkins}, {Crook}, {Cutri}, {Erdo{\v{g}}du}, {Falco}, {George}, {Hutcheson}, {Lahav}, {Mader}, {Mink}, {Martimbeau}, {Schneider}, {Skrutskie}, {Tokarz}, \& {Westover}}]{Huchra12}
{Huchra}, J.~P., {Macri}, L.~M., {Masters}, K.~L., {et~al.} 2012, \apjs, 199, 26

\bibitem[{{Huchtmeier}(1982)}]{Huchtmeier82}
{Huchtmeier}, W.~K. 1982, \aap, 110, 121

\bibitem[{{Irwin} {et~al.}(2012){Irwin}, {Beck}, {Benjamin}, {Dettmar}, {English}, {Heald}, {Henriksen}, {Johnson}, {Krause}, {Li}, {Miskolczi}, {Mora}, {Murphy}, {Oosterloo}, {Porter}, {Rand}, {Saikia}, {Schmidt}, {Strong}, {Walterbos}, {Wang}, \& {Wiegert}}]{Irwin12a}
{Irwin}, J., {Beck}, R., {Benjamin}, R.~A., {et~al.} 2012, \aj, 144, 43

\bibitem[{{Irwin} {et~al.}(2024){Irwin}, {Beck}, {Cook}, {Dettmar}, {English}, {Heesen}, {Henriksen}, {Jiang}, {Li}, {Lu}, {Mele}, {M{\"u}ller}, {Murphy}, {Porter}, {Rand}, {Skeggs}, {Stein}, {Stein}, {Stil}, {Strong}, {Walterbos}, {Wang}, {Wiegert}, \& {Yang}}]{Irwin24}
{Irwin}, J., {Beck}, R., {Cook}, T., {et~al.} 2024, Galaxies, 12, 22

\bibitem[{{Irwin} {et~al.}(2019){Irwin}, {Wiegert}, {Merritt}, {We{\.z}gowiec}, {Hunt}, {Woodfinden}, {Stein}, {Damas-Segovia}, {Li}, {Wang}, {Johnson}, {Krause}, {Dettmar}, {Im}, {Schmidt}, {Miskolczi}, {Braun}, {Saikia}, {English}, \& {Richardson}}]{Irwin19}
{Irwin}, J., {Wiegert}, T., {Merritt}, A., {et~al.} 2019, \aj, 158, 21

\bibitem[{{Irwin} {et~al.}(2015){Irwin}, {Henriksen}, {Krause}, {Wang}, {Wiegert}, {Murphy}, {Heald}, \& {Perlman}}]{Irwin15}
{Irwin}, J.~A., {Henriksen}, R.~N., {Krause}, M., {et~al.} 2015, \apj, 809, 172

\bibitem[{{Irwin} {et~al.}(2017){Irwin}, {Schmidt}, {Damas-Segovia}, {Beck}, {English}, {Heald}, {Henriksen}, {Krause}, {Li}, {Rand}, {Wang}, {Wiegert}, {Kamieneski}, {Par{\'e}}, \& {Sullivan}}]{Irwin17}
{Irwin}, J.~A., {Schmidt}, P., {Damas-Segovia}, A., {et~al.} 2017, \mnras, 464, 1333

\bibitem[{{Irwin} \& {Sofue}(1992)}]{Irwin92}
{Irwin}, J.~A. \& {Sofue}, Y. 1992, \apjl, 396, L75

\bibitem[{{Israel}(2009{\natexlab{a}})}]{Israel09a}
{Israel}, F.~P. 2009{\natexlab{a}}, \aap, 493, 525

\bibitem[{{Israel}(2009{\natexlab{b}})}]{Israel09b}
{Israel}, F.~P. 2009{\natexlab{b}}, \aap, 506, 689

\bibitem[{{Jiang} {et~al.}(2024){Jiang}, {Li}, {Gao}, {Bregman}, {Ji}, {Jiang}, {Tan}, {Wang}, {Wang}, \& {Yang}}]{Jiang24}
{Jiang}, Y., {Li}, J.-T., {Gao}, Y., {et~al.} 2024, \mnras, 528, 4160

\bibitem[{{Jim{\'e}nez-Donaire} {et~al.}(2017){Jim{\'e}nez-Donaire}, {Cormier}, {Bigiel}, {Leroy}, {Gallagher}, {Krumholz}, {Usero}, {Hughes}, {Kramer}, {Meier}, {Murphy}, {Pety}, {Schinnerer}, {Schruba}, {Schuster}, {Sliwa}, \& {Tomicic}}]{Jim17}
{Jim{\'e}nez-Donaire}, M.~J., {Cormier}, D., {Bigiel}, F., {et~al.} 2017, \apjl, 836, L29

\bibitem[{{Kennicutt}(1998)}]{Kennicutt98}
{Kennicutt}, Robert~C., J. 1998, \apj, 498, 541

\bibitem[{{Kennicutt} \& {Evans}(2012)}]{Kennicutt12}
{Kennicutt}, R.~C. \& {Evans}, N.~J. 2012, \araa, 50, 531

\bibitem[{{Krause} {et~al.}(2018){Krause}, {Irwin}, {Wiegert}, {Miskolczi}, {Damas-Segovia}, {Beck}, {Li}, {Heald}, {M{\"u}ller}, {Stein}, {Rand}, {Heesen}, {Walterbos}, {Dettmar}, {Vargas}, {English}, \& {Murphy}}]{Krause18}
{Krause}, M., {Irwin}, J., {Wiegert}, T., {et~al.} 2018, \aap, 611, A72

\bibitem[{{Krumholz} {et~al.}(2009){Krumholz}, {McKee}, \& {Tumlinson}}]{Krumholz09}
{Krumholz}, M.~R., {McKee}, C.~F., \& {Tumlinson}, J. 2009, \apj, 693, 216

\bibitem[{{Leroy} {et~al.}(2005){Leroy}, {Bolatto}, {Simon}, \& {Blitz}}]{leroy05}
{Leroy}, A., {Bolatto}, A.~D., {Simon}, J.~D., \& {Blitz}, L. 2005, \apj, 625, 763

\bibitem[{{Leroy} {et~al.}(2022){Leroy}, {Rosolowsky}, {Usero}, {Sandstrom}, {Schinnerer}, {Schruba}, {Bolatto}, {Sun}, {Barnes}, {Belfiore}, {Bigiel}, {den Brok}, {Cao}, {Chiang}, {Chevance}, {Dale}, {Eibensteiner}, {Faesi}, {Glover}, {Hughes}, {Jim{\'e}nez Donaire}, {Klessen}, {Koch}, {Kruijssen}, {Liu}, {Meidt}, {Pan}, {Pety}, {Puschnig}, {Querejeta}, {Saito}, {Sardone}, {Watkins}, {Weiss}, \& {Williams}}]{Leroy22}
{Leroy}, A.~K., {Rosolowsky}, E., {Usero}, A., {et~al.} 2022, \apj, 927, 149

\bibitem[{{Leroy} {et~al.}(2009){Leroy}, {Walter}, {Bigiel}, {Usero}, {Weiss}, {Brinks}, {de Blok}, {Kennicutt}, {Schuster}, {Kramer}, {Wiesemeyer}, \& {Roussel}}]{Leroy09}
{Leroy}, A.~K., {Walter}, F., {Bigiel}, F., {et~al.} 2009, \aj, 137, 4670

\bibitem[{{Leroy} {et~al.}(2008){Leroy}, {Walter}, {Brinks}, {Bigiel}, {de Blok}, {Madore}, \& {Thornley}}]{Leroy08}
{Leroy}, A.~K., {Walter}, F., {Brinks}, E., {et~al.} 2008, \aj, 136, 2782

\bibitem[{{Leroy} {et~al.}(2013){Leroy}, {Walter}, {Sandstrom}, {Schruba}, {Munoz-Mateos}, {Bigiel}, {Bolatto}, {Brinks}, {de Blok}, {Meidt}, {Rix}, {Rosolowsky}, {Schinnerer}, {Schuster}, \& {Usero}}]{Leroy13}
{Leroy}, A.~K., {Walter}, F., {Sandstrom}, K., {et~al.} 2013, \aj, 146, 19

\bibitem[{{Li} {et~al.}(2016{\natexlab{a}}){Li}, {Beck}, {Dettmar}, {Heald}, {Irwin}, {Johnson}, {Kepley}, {Krause}, {Murphy}, {Orlando}, {Rand}, {Strong}, {Vargas}, {Walterbos}, {Wang}, \& {Wiegert}}]{Li16a}
{Li}, J.-T., {Beck}, R., {Dettmar}, R.-J., {et~al.} 2016{\natexlab{a}}, \mnras, 456, 1723

\bibitem[{{Li} {et~al.}(2016{\natexlab{b}}){Li}, {Bregman}, {Wang}, {Crain}, \& {Anderson}}]{Li16b}
{Li}, J.-T., {Bregman}, J.~N., {Wang}, Q.~D., {Crain}, R.~A., \& {Anderson}, M.~E. 2016{\natexlab{b}}, \apj, 830, 134

\bibitem[{{Li} {et~al.}(2018){Li}, {Bregman}, {Wang}, {Crain}, \& {Anderson}}]{Li18}
{Li}, J.-T., {Bregman}, J.~N., {Wang}, Q.~D., {Crain}, R.~A., \& {Anderson}, M.~E. 2018, \apjl, 855, L24

\bibitem[{{Li} {et~al.}(2017){Li}, {Bregman}, {Wang}, {Crain}, {Anderson}, \& {Zhang}}]{Li17}
{Li}, J.-T., {Bregman}, J.~N., {Wang}, Q.~D., {et~al.} 2017, \apjs, 233, 20

\bibitem[{{Li} {et~al.}(2014){Li}, {Crain}, \& {Wang}}]{Li14}
{Li}, J.-T., {Crain}, R.~A., \& {Wang}, Q.~D. 2014, \mnras, 440, 859

\bibitem[{{Li} {et~al.}(2024){Li}, {Lu}, {Qu}, {Benjamin}, {Bregman}, {Dettmar}, {English}, {Fang}, {Irwin}, {Jiang}, {Li}, {Liu}, {Martini}, {Rand}, {Stein}, {Strong}, {Vargas}, {Wang}, {Wang}, {Wiegert}, {Xu}, \& {Yang}}]{Li24}
{Li}, J.-T., {Lu}, L.-Y., {Qu}, Z., {et~al.} 2024, \apj, 967, 78

\bibitem[{{Li} \& {Wang}(2013{\natexlab{a}})}]{Li13a}
{Li}, J.-T. \& {Wang}, Q.~D. 2013{\natexlab{a}}, \mnras, 428, 2085

\bibitem[{{Li} \& {Wang}(2013{\natexlab{b}})}]{Li13b}
{Li}, J.-T. \& {Wang}, Q.~D. 2013{\natexlab{b}}, \mnras, 435, 3071

\bibitem[{{Li} {et~al.}(2019){Li}, {Zhou}, {Jiang}, {Bregman}, \& {Gao}}]{Li19}
{Li}, J.-T., {Zhou}, P., {Jiang}, X., {Bregman}, J.~N., \& {Gao}, Y. 2019, \apj, 877, 3

\bibitem[{{Lisenfeld} {et~al.}(2011){Lisenfeld}, {Espada}, {Verdes-Montenegro}, {Kuno}, {Leon}, {Sabater}, {Sato}, {Sulentic}, {Verley}, \& {Yun}}]{Lisenfeld11}
{Lisenfeld}, U., {Espada}, D., {Verdes-Montenegro}, L., {et~al.} 2011, \aap, 534, A102

\bibitem[{{Lisenfeld} {et~al.}(2023){Lisenfeld}, {Ogle}, {Appleton}, {Jarrett}, \& {Moncada-Cuadri}}]{Lisenfeld23}
{Lisenfeld}, U., {Ogle}, P.~M., {Appleton}, P.~N., {Jarrett}, T.~H., \& {Moncada-Cuadri}, B.~M. 2023, \aap, 673, A87

\bibitem[{{Lisenfeld} {et~al.}(2019){Lisenfeld}, {Xu}, {Gao}, {Domingue}, {Cao}, {Yun}, \& {Zuo}}]{Lisenfeld19}
{Lisenfeld}, U., {Xu}, C.~K., {Gao}, Y., {et~al.} 2019, \aap, 627, A107

\bibitem[{{Lu} {et~al.}(2023){Lu}, {Li}, {Vargas}, {Beck}, {Bregman}, {Dettmar}, {English}, {Fang}, {Heald}, {Li}, {Qu}, {Rand}, {Stein}, {Wang}, {Wang}, {Wiegert}, \& {Zheng}}]{Lu23}
{Lu}, L.-Y., {Li}, J.-T., {Vargas}, C.~J., {et~al.} 2023, \mnras, 519, 6098

\bibitem[{{Lu} {et~al.}(2024){Lu}, {Li}, {Vargas}, {Fang}, {Benjamin}, {Bregman}, {Dettmar}, {English}, {Heald}, {Jiang}, {Wang}, \& {Yang}}]{Lu24}
{Lu}, L.-Y., {Li}, J.-T., {Vargas}, C.~J., {et~al.} 2024, arXiv e-prints, arXiv:2410.02347

\bibitem[{{Makarov} {et~al.}(2014){Makarov}, {Prugniel}, {Terekhova}, {Courtois}, \& {Vauglin}}]{Makarov14}
{Makarov}, D., {Prugniel}, P., {Terekhova}, N., {Courtois}, H., \& {Vauglin}, I. 2014, \aap, 570, A13

\bibitem[{{McGaugh}(2005)}]{McGaugh05}
{McGaugh}, S.~S. 2005, \apj, 632, 859

\bibitem[{{McGaugh} {et~al.}(2000){McGaugh}, {Schombert}, {Bothun}, \& {de Blok}}]{McGaugh00}
{McGaugh}, S.~S., {Schombert}, J.~M., {Bothun}, G.~D., \& {de Blok}, W.~J.~G. 2000, \apjl, 533, L99

\bibitem[{{Meier} \& {Turner}(2004)}]{Meier04}
{Meier}, D.~S. \& {Turner}, J.~L. 2004, \aj, 127, 2069

\bibitem[{{Paglione} {et~al.}(2001){Paglione}, {Wall}, {Young}, {Heyer}, {Richard}, {Goldstein}, {Kaufman}, {Nantais}, \& {Perry}}]{Paglione01}
{Paglione}, T. A.~D., {Wall}, W.~F., {Young}, J.~S., {et~al.} 2001, \apjs, 135, 183

\bibitem[{{Pan} {et~al.}(2018){Pan}, {Lin}, {Hsieh}, {Xiao}, {Gao}, {Ellison}, {Scudder}, {Barrera-Ballesteros}, {Yuan}, {Saintonge}, {Wilson}, {Hwang}, {De Looze}, {Gao}, {Ho}, {Brinks}, {Mok}, {Brown}, {Davis}, {Williams}, {Chung}, {Parsons}, {Bureau}, {Sargent}, {Chung}, {Kim}, {Liu}, {Micha{\l}owski}, \& {Tosaki}}]{Pan18}
{Pan}, H.-A., {Lin}, L., {Hsieh}, B.-C., {et~al.} 2018, \apj, 868, 132

\bibitem[{{Papadopoulos} \& {Seaquist}(1999)}]{Papadopoulos99}
{Papadopoulos}, P.~P. \& {Seaquist}, E.~R. 1999, \apj, 516, 114

\bibitem[{{Papadopoulos} {et~al.}(2012){Papadopoulos}, {van der Werf}, {Xilouris}, {Isaak}, {Gao}, \& {M{\"u}hle}}]{Papadopoulos12}
{Papadopoulos}, P.~P., {van der Werf}, P.~P., {Xilouris}, E.~M., {et~al.} 2012, \mnras, 426, 2601

\bibitem[{{Pe{\~n}aloza} {et~al.}(2018){Pe{\~n}aloza}, {Clark}, {Glover}, \& {Klessen}}]{Penaloza18}
{Pe{\~n}aloza}, C.~H., {Clark}, P.~C., {Glover}, S. C.~O., \& {Klessen}, R.~S. 2018, \mnras, 475, 1508

\bibitem[{{Pe{\~n}aloza} {et~al.}(2017){Pe{\~n}aloza}, {Clark}, {Glover}, {Shetty}, \& {Klessen}}]{Penaloza17}
{Pe{\~n}aloza}, C.~H., {Clark}, P.~C., {Glover}, S. C.~O., {Shetty}, R., \& {Klessen}, R.~S. 2017, \mnras, 465, 2277

\bibitem[{{Renaud} {et~al.}(2014){Renaud}, {Bournaud}, {Kraljic}, \& {Duc}}]{Renaud14}
{Renaud}, F., {Bournaud}, F., {Kraljic}, K., \& {Duc}, P.~A. 2014, \mnras, 442, L33

\bibitem[{{Roman-Duval} {et~al.}(2016){Roman-Duval}, {Heyer}, {Brunt}, {Clark}, {Klessen}, \& {Shetty}}]{Roman16}
{Roman-Duval}, J., {Heyer}, M., {Brunt}, C.~M., {et~al.} 2016, \apj, 818, 144

\bibitem[{{Rosolowsky} \& {Leroy}(2006)}]{Rosolowsky06}
{Rosolowsky}, E. \& {Leroy}, A. 2006, \pasp, 118, 590

\bibitem[{{Saintonge} \& {Catinella}(2022)}]{Saintonge22}
{Saintonge}, A. \& {Catinella}, B. 2022, \araa, 60, 319

\bibitem[{{Saintonge} {et~al.}(2017){Saintonge}, {Catinella}, {Tacconi}, {Kauffmann}, {Genzel}, {Cortese}, {Dav{\'e}}, {Fletcher}, {Graci{\'a}-Carpio}, {Kramer}, {Heckman}, {Janowiecki}, {Lutz}, {Rosario}, {Schiminovich}, {Schuster}, {Wang}, {Wuyts}, {Borthakur}, {Lamperti}, \& {Roberts-Borsani}}]{Saintonge17}
{Saintonge}, A., {Catinella}, B., {Tacconi}, L.~J., {et~al.} 2017, \apjs, 233, 22

\bibitem[{{Saintonge} {et~al.}(2011){Saintonge}, {Kauffmann}, {Wang}, {Kramer}, {Tacconi}, {Buchbender}, {Catinella}, {Graci{\'a}-Carpio}, {Cortese}, {Fabello}, {Fu}, {Genzel}, {Giovanelli}, {Guo}, {Haynes}, {Heckman}, {Krumholz}, {Lemonias}, {Li}, {Moran}, {Rodriguez-Fernandez}, {Schiminovich}, {Schuster}, \& {Sievers}}]{Saintonge11}
{Saintonge}, A., {Kauffmann}, G., {Wang}, J., {et~al.} 2011, \mnras, 415, 61

\bibitem[{{Salim} {et~al.}(2007){Salim}, {Rich}, {Charlot}, {Brinchmann}, {Johnson}, {Schiminovich}, {Seibert}, {Mallery}, {Heckman}, {Forster}, {Friedman}, {Martin}, {Morrissey}, {Neff}, {Small}, {Wyder}, {Bianchi}, {Donas}, {Lee}, {Madore}, {Milliard}, {Szalay}, {Welsh}, \& {Yi}}]{Salim07}
{Salim}, S., {Rich}, R.~M., {Charlot}, S., {et~al.} 2007, \apjs, 173, 267

\bibitem[{{Sandstrom} {et~al.}(2013){Sandstrom}, {Leroy}, {Walter}, {Bolatto}, {Croxall}, {Draine}, {Wilson}, {Wolfire}, {Calzetti}, {Kennicutt}, {Aniano}, {Donovan Meyer}, {Usero}, {Bigiel}, {Brinks}, {de Blok}, {Crocker}, {Dale}, {Engelbracht}, {Galametz}, {Groves}, {Hunt}, {Koda}, {Kreckel}, {Linz}, {Meidt}, {Pellegrini}, {Rix}, {Roussel}, {Schinnerer}, {Schruba}, {Schuster}, {Skibba}, {van der Laan}, {Appleton}, {Armus}, {Brandl}, {Gordon}, {Hinz}, {Krause}, {Montiel}, {Sauvage}, {Schmiedeke}, {Smith}, \& {Vigroux}}]{Sandstrom13}
{Sandstrom}, K.~M., {Leroy}, A.~K., {Walter}, F., {et~al.} 2013, \apj, 777, 5

\bibitem[{{Schmidt}(1959)}]{Schmidt59}
{Schmidt}, M. 1959, \apj, 129, 243

\bibitem[{{Shetty} {et~al.}(2011){Shetty}, {Glover}, {Dullemond}, {Ostriker}, {Harris}, \& {Klessen}}]{Shetty11b}
{Shetty}, R., {Glover}, S.~C., {Dullemond}, C.~P., {et~al.} 2011, \mnras, 415, 3253

\bibitem[{{Skrutskie} {et~al.}(2006){Skrutskie}, {Cutri}, {Stiening}, {Weinberg}, {Schneider}, {Carpenter}, {Beichman}, {Capps}, {Chester}, {Elias}, {Huchra}, {Liebert}, {Lonsdale}, {Monet}, {Price}, {Seitzer}, {Jarrett}, {Kirkpatrick}, {Gizis}, {Howard}, {Evans}, {Fowler}, {Fullmer}, {Hurt}, {Light}, {Kopan}, {Marsh}, {McCallon}, {Tam}, {Van Dyk}, \& {Wheelock}}]{Skrutskie06}
{Skrutskie}, M.~F., {Cutri}, R.~M., {Stiening}, R., {et~al.} 2006, \aj, 131, 1163

\bibitem[{{Solomon} {et~al.}(1987){Solomon}, {Rivolo}, {Barrett}, \& {Yahil}}]{Solomon87}
{Solomon}, P.~M., {Rivolo}, A.~R., {Barrett}, J., \& {Yahil}, A. 1987, \apj, 319, 730

\bibitem[{{Solomon} {et~al.}(1979){Solomon}, {Scoville}, \& {Sanders}}]{Solomon79}
{Solomon}, P.~M., {Scoville}, N.~Z., \& {Sanders}, D.~B. 1979, \apjl, 232, L89

\bibitem[{{Strong} \& {Mattox}(1996)}]{Strong96}
{Strong}, A.~W. \& {Mattox}, J.~R. 1996, \aap, 308, L21

\bibitem[{{Sutter} \& {Fadda}(2022)}]{Sutter22}
{Sutter}, J. \& {Fadda}, D. 2022, \apj, 941, 47

\bibitem[{{Tacconi} {et~al.}(2020){Tacconi}, {Genzel}, \& {Sternberg}}]{Tacconi20}
{Tacconi}, L.~J., {Genzel}, R., \& {Sternberg}, A. 2020, \araa, 58, 157

\bibitem[{{Tan} {et~al.}(2011){Tan}, {Gao}, {Zhang}, \& {Xia}}]{Tan11}
{Tan}, Q.-H., {Gao}, Y., {Zhang}, Z.-Y., \& {Xia}, X.-Y. 2011, Research in Astronomy and Astrophysics, 11, 787

\bibitem[{{Tremonti} {et~al.}(2004){Tremonti}, {Heckman}, {Kauffmann}, {Brinchmann}, {Charlot}, {White}, {Seibert}, {Peng}, {Schlegel}, {Uomoto}, {Fukugita}, \& {Brinkmann}}]{Tremonti04}
{Tremonti}, C.~A., {Heckman}, T.~M., {Kauffmann}, G., {et~al.} 2004, \apj, 613, 898

\bibitem[{{Truran}(1977)}]{Truran77}
{Truran}, J.~W. 1977, in Astrophysics and Space Science Library, Vol.~67, CNO Isotopes in Astrophysics, ed. J.~{Audouze}, 145

\bibitem[{{Vargas} {et~al.}(2018){Vargas}, {Mora-Partiarroyo}, {Schmidt}, {Rand}, {Stein}, {Walterbos}, {Wang}, {Basu}, {Patterson}, {Kepley}, {Beck}, {Irwin}, {Heald}, {Li}, \& {Wiegert}}]{Vargas18}
{Vargas}, C.~J., {Mora-Partiarroyo}, S.~C., {Schmidt}, P., {et~al.} 2018, \apj, 853, 128

\bibitem[{{Vargas} {et~al.}(2019){Vargas}, {Walterbos}, {Rand}, {Stil}, {Krause}, {Li}, {Irwin}, \& {Dettmar}}]{Vargas19}
{Vargas}, C.~J., {Walterbos}, R. A.~M., {Rand}, R.~J., {et~al.} 2019, \apj, 881, 26

\bibitem[{{Vila-Vilaro} {et~al.}(2015){Vila-Vilaro}, {Cepa}, \& {Zabludoff}}]{vila15}
{Vila-Vilaro}, B., {Cepa}, J., \& {Zabludoff}, A. 2015, \apjs, 218, 28

\bibitem[{{Walter} {et~al.}(2008){Walter}, {Brinks}, {de Blok}, {Bigiel}, {Kennicutt}, {Thornley}, \& {Leroy}}]{Walter08}
{Walter}, F., {Brinks}, E., {de Blok}, W.~J.~G., {et~al.} 2008, \aj, 136, 2563

\bibitem[{{Wang} {et~al.}(2024){Wang}, {Lin}, {Yang}, {Staveley-Smith}, {Walter}, {Wang}, {Wang}, {Battisti}, {Catinella}, {Chen}, {Cortese}, {Fisher}, {Ho}, {Ji}, {Jiang}, {Kauffmann}, {Kong}, {Liu}, {Shao}, {Wang}, {Wang}, \& {Wang}}]{Wang24}
{Wang}, J., {Lin}, X., {Yang}, D., {et~al.} 2024, arXiv e-prints, arXiv:2404.09422

\bibitem[{{Wang} {et~al.}(2023){Wang}, {Yang}, {Oh}, {Staveley-Smith}, {Wang}, {Wang}, {Hess}, {Ho}, {Hou}, {Jing}, {Kamphuis}, {Li}, {Lin}, {Liu}, {Shao}, {Wang}, \& {Zhu}}]{Wang23}
{Wang}, J., {Yang}, D., {Oh}, S.~H., {et~al.} 2023, \apj, 944, 102

\bibitem[{{Werk} {et~al.}(2014){Werk}, {Prochaska}, {Tumlinson}, {Peeples}, {Tripp}, {Fox}, {Lehner}, {Thom}, {O'Meara}, {Ford}, {Bordoloi}, {Katz}, {Tejos}, {Oppenheimer}, {Dav{\'e}}, \& {Weinberg}}]{Werk14}
{Werk}, J.~K., {Prochaska}, J.~X., {Tumlinson}, J., {et~al.} 2014, \apj, 792, 8

\bibitem[{{Wiegert} {et~al.}(2015){Wiegert}, {Irwin}, {Miskolczi}, {Schmidt}, {Mora}, {Damas-Segovia}, {Stein}, {English}, {Rand}, {Santistevan}, {Walterbos}, {Krause}, {Beck}, {Dettmar}, {Kepley}, {Wezgowiec}, {Wang}, {Heald}, {Li}, {MacGregor}, {Johnson}, {Strong}, {DeSouza}, \& {Porter}}]{Wiegert15}
{Wiegert}, T., {Irwin}, J., {Miskolczi}, A., {et~al.} 2015, \aj, 150, 81

\bibitem[{{Wilson} {et~al.}(2013){Wilson}, {Rohlfs}, \& {H{\"u}ttemeister}}]{Wilson13}
{Wilson}, T.~L., {Rohlfs}, K., \& {H{\"u}ttemeister}, S. 2013, {Tools of Radio Astronomy}

\bibitem[{{Wong} \& {Blitz}(2002)}]{Wong02}
{Wong}, T. \& {Blitz}, L. 2002, \apj, 569, 157

\bibitem[{{Xu} {et~al.}(2024){Xu}, {Wang}, {Li}, \& {Chen}}]{Xu24}
{Xu}, X., {Wang}, J., {Li}, Z., \& {Chen}, Y. 2024, \apj, 971, 165

\bibitem[{{Yajima} {et~al.}(2021){Yajima}, {Sorai}, {Miyamoto}, {Muraoka}, {Kuno}, {Kaneko}, {Takeuchi}, {Yasuda}, {Tanaka}, {Morokuma-Matsui}, \& {Kobayashi}}]{Yajima21}
{Yajima}, Y., {Sorai}, K., {Miyamoto}, Y., {et~al.} 2021, \pasj, 73, 257

\bibitem[{{Yao} {et~al.}(2003){Yao}, {Seaquist}, {Kuno}, \& {Dunne}}]{Yao03}
{Yao}, L., {Seaquist}, E.~R., {Kuno}, N., \& {Dunne}, L. 2003, \apj, 588, 771

\bibitem[{{Yim} {et~al.}(2011){Yim}, {Wong}, {Howk}, \& {van der Hulst}}]{Yim11}
{Yim}, K., {Wong}, T., {Howk}, J.~C., \& {van der Hulst}, J.~M. 2011, \aj, 141, 48

\bibitem[{{Young} \& {Knezek}(1989)}]{Young89}
{Young}, J.~S. \& {Knezek}, P.~M. 1989, \apjl, 347, L55

\bibitem[{{Zanchettin} {et~al.}(2023){Zanchettin}, {Feruglio}, {Massardi}, {Lapi}, {Bischetti}, {Cantalupo}, {Fiore}, {Bongiorno}, {Malizia}, {Marinucci}, {Molina}, {Piconcelli}, {Tombesi}, {Travascio}, {Tozzi}, \& {Tripodi}}]{Zanchettin23}
{Zanchettin}, M.~V., {Feruglio}, C., {Massardi}, M., {et~al.} 2023, \aap, 679, A88

\bibitem[{{Zheng} {et~al.}(2022){Zheng}, {Wang}, {Irwin}, {English}, {Ma}, {Wang}, {Wang}, {Wang}, {Krause}, {Randriamampandry}, {Li}, \& {Beck}}]{Zheng22}
{Zheng}, Y., {Wang}, J., {Irwin}, J., {et~al.} 2022, \mnras, 513, 1329

\end{thebibliography}
\bibliographystyle{aasjournal}

\end{document}